\documentclass[12pt]{article}
\usepackage{amsmath,amssymb,amsthm,enumerate,graphicx}
\usepackage{ifthen,latexsym,syntonly}
\usepackage{setspace}
\usepackage{mathabx}
\usepackage{color}
\usepackage[round,comma,authoryear]{natbib}   % for natbib
\usepackage{float}
\usepackage{epstopdf}
\usepackage{mathabx}
\usepackage{soul}
\usepackage{pstricks}
\usepackage[hidelinks]{hyperref}
\usepackage{placeins}

\usepackage{color}
\allowdisplaybreaks[3]
\usepackage{calc,xspace}
\definecolor{wacky}{rgb}{1,0,1}

\newtheorem{theorem}{Theorem}[section]
\newtheorem{remark}[theorem]{Remark}

\newtheorem{Restriction}[theorem]{Restriction}

\newtheorem{lemma}[theorem]{Lemma}
\newtheorem{problem}[theorem]{Problem}
\newtheorem{proposition}[theorem]{Proposition}

\hypersetup{colorlinks=false}

\title{New Formulations of Ambiguous Volatility with an Application to Optimal Dynamic Contracting\footnote{I acknowledge helpful comments and suggestions from Anne Balter, Hui Chen, Sharada Dharmasankar, Leonid Kogan, Andrey Malenko, Jianjun Miao, Jian Sun, Xiangyu Zhang, three anonymous referees, and participants at the MIT Finance lunch seminar and the Becker Friedman Institute mini-conference on Ambiguity and Robustness. I am especially grateful to Lars Hansen, Andrey Malenko, and Tom Sargent (editor) for their support and feedback which greatly improved the paper. }}

\author{Peter G. Hansen\footnote{Sloan School of Management, Massachusetts Institute of Technology, 100 Main St, Cambridge, MA 02142. Email: \href{mailto:pghansen@mit.edu}{\url{pghansen@mit.edu}} }  }

\date{ \today }

\begin{document}

\maketitle

\begin{abstract}
I introduce novel preference formulations which capture aversion to ambiguity about unknown and potentially time-varying volatility. I compare these preferences with Gilboa and Schmeidler's maxmin expected utility as well as variational formulations of ambiguity aversion. The impact of ambiguity aversion is illustrated in a simple static model of portfolio choice, as well as a dynamic model of optimal contracting under repeated moral hazard. Implications for investor beliefs, optimal design of corporate securities, and asset pricing are explored. \\

%I study a continuous-time principal-agent model with hidden action in which the principal and the agent have ambiguous beliefs about the volatility of the project cash flows. I describe a novel formulation that captures uncertainty about the underlying volatility process show how it affects the optimal contract. Ambiguity aversion generates endogenous belief heterogeneity between the principal and the agent. Under the optimal contract, the agent always trusts the benchmark probability model, while the principal forms expectations as if volatility is strictly higher and state-dependent. Additionally, I show ambiguity aversion generates asset pricing implications for the implied financial securities. 

\end{abstract}

%%%% Things to add
%%% go through sections (1) and (2) and add word ``adversarial nature''
%%% cite Peng in introduction
%%% Check off everything in referee reports. 
%%% Say some words somewhere about financial slack (mayoe be this goes in conclusion? cite Wang paper)

\clearpage

\section{Introduction}

There is ample evidence that time-varying stochastic volatility exists and has important effects on real macroeconomic variables and is important in understanding empirical features of financial markets. The empirical evidence suggests that volatility follows complicated nonlinear dynamics, which often leads model builders to write down complicated parametric models of the evolution of volatility as well as its correlation with other economic quantities of interest. An obvious concern with this approach is whether it is possible for economic agents to learn or estimate these models precisely, or through repeated observation develop confidence that a particular parametric model is correct. While one may argue that this concern is unwarranted in financial markets with high-frequency observations which make volatility effectively observable,\footnote{Even in high-frequency settings, direct statistical measurement of volatility may be significantly confounded by microstructure and liquidity effects. See for example \cite{zhang2005tale}. } there is no convincing reason to dismiss these concerns as they pertain to real variables which are often observed at low frequencies. 

Motivated by these concerns, I propose new preferences that capture nonparametric model uncertainty about an unknown, possibly time-varying volatility process. These preference formulations build on existing models of ambiguity aversion, notably being a special case of the variational preferences proposed by \cite{MMR2006a}.\footnote{See \cite{maccheroni2006dynamic} for axiomatic approaches to dynamic variational preferences.} These preferences, which I call \emph{moment-constrained variational preferences}, are first formulated in a static setting of decision-making under uncertainty.  I illustrate the impact of these static preferences in a simple portfolio choice problem, where I show how the degree of ambiguity aversion affects both the implied worst-case beliefs of the investor and the optimal portfolio weight. Then, I explore dynamic counterparts to these preferences and derive a continuous-time limit in which the decision-maker is uncertain about an unknown, time-varying volatility process. The impact of these preferences is then illustrated in a model of repeated moral hazard based on papers by \cite{demarzo2006optimal} and \cite{biais2007dynamic}. Using this model, I explore the implications of ambiguity aversion for optimal security design and asset prices, and compare and contrast the implications of dynamic variational preferences with those of $G$-expectations. 

Ambiguity aversion leads the principal to design a contract that is robustly optimal given uncertainty about the volatility process. Under the optimal contract, belief heterogeneity emerges between the principal and the agent. The agent trusts the benchmark volatility model, whereas the principal forms expectations as if volatility is strictly higher and state-dependent. As in \cite{demarzo2006optimal}, the optimal contract can be interpreted as featuring a line of credit between the principal and the agent. I show how ambiguity aversion increases the optimal credit limit, while reducing the reliance on long-term debt. This is important since credit lines are a commonly used corporate security. Additionally, I derive asset pricing implications of volatility ambiguity under the optimal contract. 

%%%%%
%Say something about Prat and Jovanovic
%\cite{prat2014dynamic}
%Sung...?

%Cite Gilboa-Schmeidler, Hansen-Sargent, Hansen Nobel lecture, Bergemann-Morris, Wolitzky, Wilson?

%Why ambiguous volatility?
%Tons of evidence for stochastic volatility in real world
%Leads many to complicated stochastic volatility models
%Is this plausible? Can agents really estimate / fully understand this complexity
%Rather in my model agents think about a ``robust'' stochastic volatility model
%Cts time implication of observable volatility seems like special result, violated in any discrete time model
%Say something about ellsberg paradox...?

\subsection{Related literature}

This paper builds on a large literature on ambiguity aversion and model uncertainty in the context of economic decisions. The preference formulations I introduce fit within the framework of variational preferences introduced by \cite{MMR2006a} which nests the maxmin expected utility of \cite{Gilboa_Schmeidler} and the ``multiplier'' formulation of ambiguity due to \cite{hansensargent:2001}. Dynamic models of ambiguity and robustness can be broadly thought of as belonging to one of three categories, namely the ``recursive multiple priors'' model proposed by  \cite{epstein2003recursive}, the ``recursive smooth ambiguity'' model proposed by \cite{klibanoff2009recursive}, and the ``dynamic variational preferences'' model described in \cite{maccheroni2006dynamic} and \cite{hansen2018structured} as a generalization of the ``multiplier preferences'' introduced by \cite{hansensargent:2001}. My paper adds to this literature by proposing a new form of preferences that captures ambiguity or uncertainty about volatility in continuous time. To my knowledge, the only other model of volatility ambiguity is the ``G-expectations'' model of \cite{peng2007g}, which can be interpreted as a recursive multiple priors model. The continuous-time preferences developed in this paper can be thought of as a particular continuous-time limit of the discrete-time preferences of \cite{maccheroni2006dynamic} which nest the G-expectations model.

Models of financial contracting typically assume that all economic actors fully understand the model environment, and that such understanding is common knowledge. This is similar to (but stronger than) an assumption of rational expectations, and has been criticized as overly restrictive in models with strategic interaction by \cite{harsanyi1967games}, \cite{wilson1987}, \cite{bergemann2005robust}, \cite{woodford2010robustly}, \cite{hansen2012three}, and others. This paper attempts to relax the assumption that economic actors fully understand their model environment and study the corresponding effect on financial contracting. In particular, I study a long-term contracting problem where economic actors have ambiguous beliefs about the possibly time-varying volatility of future cash flows. 

My paper builds on the large literature studying models of long-term financial contracting. \cite{demarzo2007optimal},  \cite{demarzo2006optimal}, and \cite{biais2007dynamic} show that in stationary environments with risk-neutral economic agents, the optimal long-term financial contract can be implemented by an interpretable capital structure. I build on these papers by introducing uncertainty about the volatility of the cash flow process and study how this affects the optimal contract. As with many of these papers, I rely on the martingale approach to dynamic contracting problems developed by \cite{sannikov2008continuous} and \cite{williams2008dynamic}. 

Particularly relevant are papers that take robust approaches to incentive problems, such as \cite{bergemann2005robust}, \cite{carroll2015robustness},  \cite{zhu2016renegotiation}, and \cite{malenko2018asymmetric}. The closest paper to this one is \cite{miao2016robust} who characterize the optimal contract in continuous time when the principal faces ambiguity about expected cash flows. As I will demonstrate, my model produces substantially different optimal security design yet has qualitatively similar asset pricing implications. \cite{szydlowski2012ambiguity} and \cite{prat2014dynamic} study related problems where the principal is uncertain about the details of the agency problem. \cite{adrian2009disagreement} characterize optimal contracting when the principal and the agent disagree about the underlying dynamics of the cash flow process and both learn through time. By focusing on uncertainty about second moments, my paper is similar in spirit to \cite{wolitzky2016mechanism} who studies a static mechanism design problem.

\subsection{Outline}
The outline of this paper is as follows. Section 2 defines \emph{moment-constrained variational preferences} in the context of static decision problems and then illustrates the impact of these preferences in a static model of portfolio choice under quadratic utility. Section 3 extends these preferences to dynamic problems in a discrete-time setting and explores an interesting continuous-time limit under which the more general ambiguity about the probability distribution of state evolution reduces to ambiguity about an unknown stochastic volatility process. I compare the continuous-time limit to the $G$-expectations model of \cite{peng2007g}. Section 4 applies the continuous-time preferences to a model of optimal contracting under repeated moral hazard and illustrates the implications of ambiguous volatility for security design and asset pricing. Section 5 concludes.

\section{Static preferences}

Let $P$ denote the decision-maker's benchmark probability measure, and let $\mathbb{E}_P[ \cdot ]$ denote the expectation operator under $P$. Let $a \in \mathcal{A}$ denote the action of the decision-maker and let $\epsilon$ denote a vector of payoff-relevant shocks. I would like to capture the notion that the decision-maker is uncertain about the entire distribution $P$, but is certain about certain moments or functionals of the distribution $P$. More precisely, let $g(\cdot )$ be any function such that $\mathbb{E}_P[ g(\epsilon) ] = 0$. The decision-maker allows for absolutely continuous probabilistic distortions of $P$, which I parameterize by likelihood ratio random variables $M$ which satisfy $M \geq 0$ with $P$-probability 1 and $\mathbb{E}_P[M] = 1$. The decision-maker's certainty about the random variable $g(\epsilon)$ being mean zero is captured by restricting the set of likelihood ratios to those that satisfy $\mathbb{E}_P [ M g(\epsilon) ] = 0$. 

Define investor utility $V(a;P,\theta)$ as
\begin{equation}
\label{pref1}
V(a;P,\theta) = \underset{M \geq 0, \\ \mathbb{E}_P[M] = 1}{\inf}  \mathbb{E}_P[M U(a,\epsilon)] + \theta \Phi(M) 
\end{equation}
where
\[
\Phi(M) = 
\begin{cases} 
\mathbb{E}_P[ c(M) ] &\text{ if } \mathbb{E}_P[M g(\epsilon) ] = 0 \\ 
\infty & \text{otherwise} 
\end{cases}
\]
where $c(\cdot )$ is a convex function with $c(1) = 0$.\footnote{Note that the penalization function $\Phi(\cdot)$ is equivalent to an $f$-divergence on the set of $M$'s which satisfy the moment restriction. } If the infimum in equation \eqref{pref1} is attained, I refer to the $M$ that attains it as the \emph{worst-case} belief distortion. 
Note that the penalty function is infinite if the moment restriction $\mathbb{E}_P \left[ M g(\epsilon ) \right] = 0$ is violated. This captures the notion that the investor is certain about this aspect of the distribution, because the worst-case belief distortion $M$ will never violate the moment restriction. Since the penalty function $\Phi(\cdot )$ is convex in $M$, these preferences fit into the variational preference framework axiomatized by \cite{MMR2006a}. I therefore refer to preferences defined by \eqref{pref1} as moment-constrained variational preferences. 

The decision-maker's problem can then expressed concisely as
\[
\max_{a \in \mathcal{A}} V(a;P,\theta). 
\]

\subsection{Relative entropy penalization}
A particularly tractable choice of divergence function $c(\cdot )$ is given by $c(M) = M \log M$.\footnote{This corresponds to an $f$-divergence known as relative entropy or Kullback-Leibler divergence, and has numerous statistical interpretations. Certain members of the larger family of power divergences proposed by \cite{cressie1984multinomial}  including Hellinger and logarithmic divergences, as well as the related notion of Chernoff entropy, can lead to degenerate solutions to \eqref{pref1}. See \cite{chenrobust} for further discussion. } This divergence gives particularly tractable expressions for the worst-case likelihood ratio $M^*$. In particular, by standard convex duality arguments, it is possible to show that $M^*$ has the following exponential tilting form
\begin{equation}
\label{et}
M^*(a,\theta) = \frac{ \exp \left( - \theta^{-1} U(a,\epsilon) - \lambda(a,\theta)g(\epsilon)  \right) }{\mathbb{E}_P \left[ \exp \left( - \theta^{-1} U(a,\epsilon) - \lambda(a,\theta)g(\epsilon)  \right)  \right] }
\end{equation}
where the Lagrange multiplier $\lambda(a,\theta)$ is chosen to satisfy the moment restriction 
\[
\mathbb{E}_P \left[ M^*(a,\theta) g(\epsilon)  \right] = 0.
\] 

It is worth comparing the expression for the likelihood ratio in equation \eqref{et} to the corresponding expression with the constraint $\mathbb{E}_P[M g(\epsilon)] = 0$ omitted. This would correspond to ``multiplier'' preferences. In this case, the corresponding worst-case likelihood ratio would be given by
\begin{equation}
\label{mult}
M^*(a,\theta) = \frac{\exp (- \theta^{-1} U(a,\epsilon) )}{\mathbb{E}_P \left[ \exp ( - \theta^{-1} U(a,\epsilon) ) \right] }. 
\end{equation}
In equation \eqref{mult}, we see that the worst-case likelihood ratio distorts probabilities towards states in which the decision-maker's utility is low. Such a distortion will generally violate the moment condition. By contrast, the likelihood ratio in equation \eqref{et} must distort probabilities in such a way that the moment restriction remains satisfied. In particular, for concave utility functions, this intuitively corresponds to a likelihood ratio that increases the dispersion of $\epsilon$ by overweighting states in which $\epsilon$ takes values that are far from its mean. This will be seen in the example presented in the next section. 

%%%% RELATIVE ENTROPY AND EXPONENTIAL TILTING
%% cite MMR earlier and more often
%% also cite Hansen and Sargent (do comparison with multiplier preferences)
%% Also also cite Chen et al. for problems with other Cressie-Read divergences

\subsection{Example: Portfolio choice under quadratic utility}

I illustrate the impact of these preferences in a static portfolio choice problem. Consider an investor with quadratic utility
\[
U\left(\widetilde{W}\right) = - \frac{1}{2} \left( \widetilde W - b\right)^2
\]
over period-1 wealth $\widetilde{W}$ where $b$ is the investor's bliss-point wealth level. Investor has initial wealth $W_0$ which can be invested at a gross risk-free rate $R_f$ or a vector of risky assets with excess return vector $\widetilde R$.

While quadratic utility has the well-known and arguably undesirable feature that the investor's utility can be decreasing in wealth,  I include this example for two reasons. First, it leads to quasi-analytic solutions which facilitate intuition. Second, in continuous-time diffusion limits, local quadratic approximations become exact. Therefore, much of the intuition obtained from the static linear-quadratic Gaussian model examined in this section will carry over to more general continuous-time diffusion models studied later. 

\subsubsection{Benchmark: No ambiguity}
Under the investor's subjective probability measure $P$, the vector of excess returns $R$ is distributed as $R \sim \text{Normal}(\mu,\Sigma)$. Formally, the investor's portfolio optimization problem can be written as
\begin{align*}
\max_{\phi \in \mathbb{R}^k} \ &\mathbb{E}_P \left[ U\left( \widetilde{W} \right)  \right] \\
\text{s.t. } & \widetilde{W} = W_0 R_f + \phi' \widetilde{R}.
\end{align*}
One can verify that the optimal portfolio weight vector $\phi^*$ is given by
\[
\phi^* = \left[ \mu \mu' + \Sigma \right]^{-1} \mu \left( b - W_0 R_f \right). 
\]
\subsubsection{Portfolio choice with ambiguity}
Consider the same portfolio choice problem as before, but now the investor treats their subjective probability measure $P$ under which $\widetilde{R} \sim \text{Normal}(\mu,\Sigma)$ as an approximation. They are willing to entertain other probability measures as possible, but treat the expected return vector $\mathbb{E}[\widetilde{R}] = \mu$ as certain. Then we can model the investor's preferences as
\[
V(\phi; P,\theta) 
= \inf_{M \geq 0, \mathbb{E}_P[M] = 1} \mathbb{E}_P \left[ M U\left(\widetilde{W} \right) \right] 
+ \theta \Phi(M) 
\]
where
\[
\Phi(M) = \begin{cases} 
\mathbb{E}_P \left[ M \log M \right] & \text{ if } \mathbb{E}_P \left[ M \left( \widetilde{R}  - \mu \right) \right] = 0 \\
\infty & \text{ otherwise.}
\end{cases}
\]
The investor's portfolio choice problem is then
\[
\max_{\phi \in \mathbb{R}^k} \ V(\phi; P,\theta).
\]
As in the expected utility case, the investor's problem can be solved in closed form. First, for fixed $\phi \in \mathbb{R}^k$ we can solve the infimization problem for $V(\phi;P,\theta)$. Using standard duality results, it can be shown that the worst-case $M$ is unique and given by
\[
M\left(\widetilde{R};\phi \right) = \frac{ \exp \left( - \frac{1}{2 \theta } \left( \widetilde{R} - \mu \right)' \left[ - \phi \phi' \right] \left( \widetilde{R} - \mu \right) \right)}{ \mathbb{E}_P \left[ \exp \left( - \frac{1}{2\theta} \left( \widetilde{R} - \mu \right)' \left[ -  \phi \phi' \right] \left( \widetilde{R} - \mu \right) \right)  \right] }
\]
for choices of $\phi$ for which the objective function is finite.\footnote{This need not be the case. For some choices for $\phi$ the adversarial nature's choice of $M$ can make the investor's utility arbitrarily negative. } This shows that for any $\phi$ the resulting penalized worst-case probability distribution is also a multivariate normal distribution. As expected, we have $\mathbb{E}_P \left[ M\left(\widetilde{R};\phi \right) \right] = 1$ and $\mathbb{E}_P \left[  \left(\widetilde{R};\phi \right) \widetilde{R} \right]  = \mu$, so there is no mean distortion. Additionally, we can see that the worst-case $M$ distorts the variance-covariance matrix of the returns from $\Sigma$ to $\left[ \Sigma^{-1} - \theta^{-1} \phi \phi' \right]^{-1}$.

Due to the regularity of the problem, it can be shown that the orders of minimization and maximization can be exchanged at the optimal $M^*$ and $\theta^*$. Exchanging the order of maximization and minimization, we can solve for the optimal $\phi$ as a function of $M$. We obtain easily that
\[
\phi(M) = \mathbb{E} \left[ M \widetilde{R} \widetilde{R}' \right]^{-1} \mu (b - W_0 R_f).
\]
Next observe that at the optimal $M^*$ must satisfy
\[
M^* = M \left( \widetilde{R} ; \phi(M^*) \right).
\]
Write $S = \mathbb{E} \left[ M^* \widetilde{R} \widetilde{R}' \right]$. By our previous observation, we must have that $S$ is a positive-definite solution of the equation
\[
S - \mu \mu' = \left[ \Sigma^{-1} - \theta^{-1} (b - W_0 R_f) \mu' S^{-2} \mu (b - W_0 R_f) \right]^{-1}.
\]
where this equation can be obtained by simply writing the penalized worst-case distorted variance of $\widetilde{R}$ in two ways. Unfortunately, this equation is a third-order matrix polynomial in $S$ so we cannot easily express $S$ in closed form. Nonetheless, we know that under the penalized worst-case distribution of $\widetilde{R}$ we have
\[
\widetilde{R} \sim \text{Normal} \left(\mu, S - \mu \mu' \right).
\]
and that
\[
\phi^*(\theta) = S^{-1} \mu (b - W_0 R_f). 
\]
%It can be shown that
%\begin{align*}
%V(\phi;P,\theta) 
%= &- \frac{1}{2} \left[ (W_0 R_f - b)^2 + 2 \phi' \mu (W_0 R_f - b ) + \phi' \left( \mu \mu' + \left( \Sigma^{-1} - \theta^{-1} \phi \phi'  \right)^{-1} \right)\phi    \right] \\
%&+ \frac{1}{2} \left[ \text{tr} \left( \Sigma^{-1} \left( \Sigma^{-1} - \theta^{-1} \phi \phi' \right)^{-1} \right) - k + \log \left( \text{det} \left( \Sigma  \left( \Sigma^{-1} - \theta^{-1} \phi \phi' \right)  \right)  \right)  \right] 
%\end{align*}
%It follows directly that as $\theta \to \infty$ we recover the expected utility solution. 

\subsubsection{Scalar risky asset}
\label{subsec:scalar}
To facilitate intuition further, I focus on the case where the risky return $\tilde R$ is a scalar random variable, so that the portfolio weight $\phi$ is also a scalar. Assume that $\tilde R$ has mean $\mu$ and benchmark variance $\sigma^2$. Under expected utility, the optimal portfolio weight is
\[
\phi^* = \frac{1}{\mu^2 + \sigma^2}\mu (b - W_0 R_f).
\]
To characterize the solution under ambiguity aversion, it is helpful to define the scalar $s(\theta) = \mathbb{E}[M^* \widetilde{R}^2]$. Note that the optimal portfolio weight $\phi^*(\theta)$ can be written in terms of $s(\theta)$ as
\[
\phi^*(\theta) = \frac{1}{s} \mu (b - W_0 R_f).
\]

By the arguments in the previous section, we can write $s(\theta)$ as the solution of the minimization problem of the adversarial nature, which simplifies to
\begin{equation}
\label{smin}
\min_s - \frac{1}{2} (b - W_0 R_f)^2 + \frac{1}{2} \mu^2 (b - W_0 R_f)^2 \frac{1}{s} + \frac{\theta}{2} \left[ \frac{s - \mu^2}{\sigma^2} - 1 - \log \left( \frac{s - \mu^2}{\sigma^2} \right)  \right] . 
\end{equation}
Note that the objective function in \eqref{smin} is strictly convex in $s$, so $s(\theta)$ is uniquely defined. Additionally, it is easy to see that for $\theta > 0$, we must have $s(\theta) > \mu^2 + \sigma^2$. It follows immediately that $\phi^*(\theta) \in (0,\phi^*)$. 

%Under ambiguity aversion, the results of the previous section give the following expression for $s = \mathbb{E}[M^* \widetilde{R}^2]$
%\[
%s - \mu^2 = \left(  \frac{1}{\sigma^2} - \theta^{-1} (b - W_0 R_f)^2 \mu^2 \frac{1}{s^2} \right)^{-1}
%\]

While it is possible to obtain closed-form expressions for $s(\theta)$ and correspondingly $\phi^*(\theta)$ as roots of the cubic polynomial under appropriate inequality restrictions, these expressions are complicated and convey little intuition. Instead I characterize the comparative statics of these quantities in propositions \ref{s1} and \ref{s2} below. The solutions $s(\theta)$ and $\phi^*(\theta)$ are characterized more explicitly in the appendix in terms of the unique positive root of a particular cubic polynomial. 

%Our expression for $s$ can be re-written as a cubic polynomial in $s$
%\[
%0 = \frac{s^3}{\sigma^2} - \left( 1 + \frac{\mu^2}{\sigma^2}  \right) s^2 - \theta^{-1} \left( b - W_0 R_f \right)^2 \mu^2 s + \theta^{-1} \left(  b - W_0 R_f  \right)^2 \mu^4 
%\]

\begin{proposition}
\label{s1}
Assume that $b > W_0 R_f$ and $\mu > 0$. Then the optimal portfolio weight $\phi^*(\theta)$ has the following properties:
\begin{enumerate}[(i)]
\item $\phi^*(\theta) > 0$. 
\item $\phi^*(\theta)$ is strictly increasing in $\theta$. 
\item As $\theta \to \infty$ we have $\phi^*(\theta) \to \phi^*$. 
\item As $\theta \to 0$ we have $\phi^*(\theta) \to 0$. 
\end{enumerate}
\end{proposition}

\begin{figure}[h!]
\centering
\includegraphics[scale=0.5]{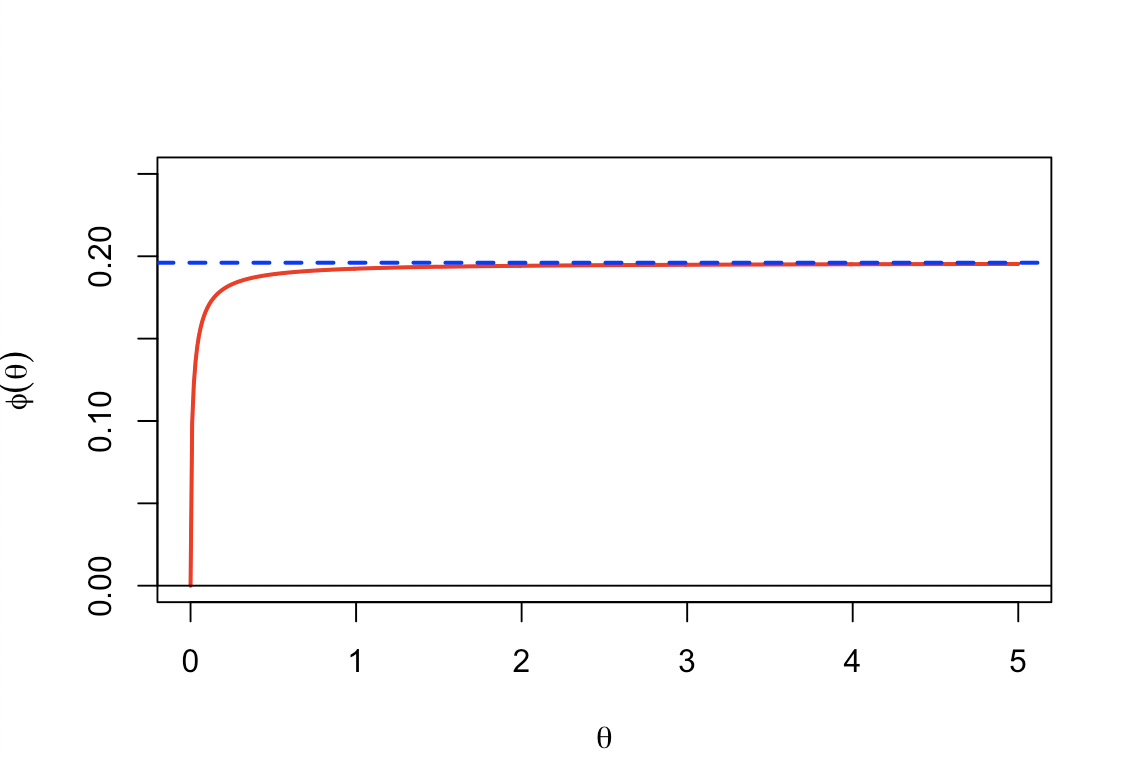}
\caption{$\phi^*(\theta)$ as a function of $\theta$ plotted in red. Horizontal line at $\phi^*$ shown in dashed blue.  Parameter values are $b = 1$, $W_0 = 0$, $\mu = 0.1$, $\sigma^2 = 0.5$. Note that $\phi^*(0) = 0$. }
\label{figs1}
\end{figure}

The results of proposition \ref{s1} are illustrated in figure \ref{figs1}. Result $(i)$ shows that the investor will always invest a strictly positive amount of their wealth in the risky asset. This makes intuitive sense because $\mu > 0$. Result $(ii)$ shows that the amount the investor's portfolio weight on the risky asset is increasing in their model confidence $\theta$ or equivalently decreasing in their ambiguity aversion $1/\theta$. Result $(iii)$ shows that as the investor's model confidence becomes infinite, their optimal portfolio weight converges to the optimal portfolio weight without ambiguity aversion. Result $(iv)$ shows that as the investor becomes infinitely ambiguity averse, they will invest none of their wealth in the risky asset. This is because they perceive the risky asset as becoming infinitely risky. 

Write $\nu^2(\theta)$ as the ratio of the worst-case volatility to the benchmark volatility, formally
\[
\nu^2(\theta) \equiv \frac{s(\theta) - \mu^2}{\sigma^2} =  \frac{\mathbb{E}[ M^*(\theta) \widetilde{R}^2 ] - \mu^2}{\mathbb{E}[\widetilde{R}^2] - \mu^2}
\]
Then we have the following continuity result:
\begin{proposition}
\label{s2}
Under the conditions of proposition \ref{s1} we have the following limits:
\begin{enumerate}[(i)]
\item As $\theta \to \infty$ we have $\nu^2(\theta) \to 1$. 
\item As $\theta \to 0$ we have $\nu^2(\theta) \to \infty$. 
\end{enumerate}
\end{proposition}

\begin{figure}[h!]
\label{figs2}
\centering
\includegraphics[scale = 0.5]{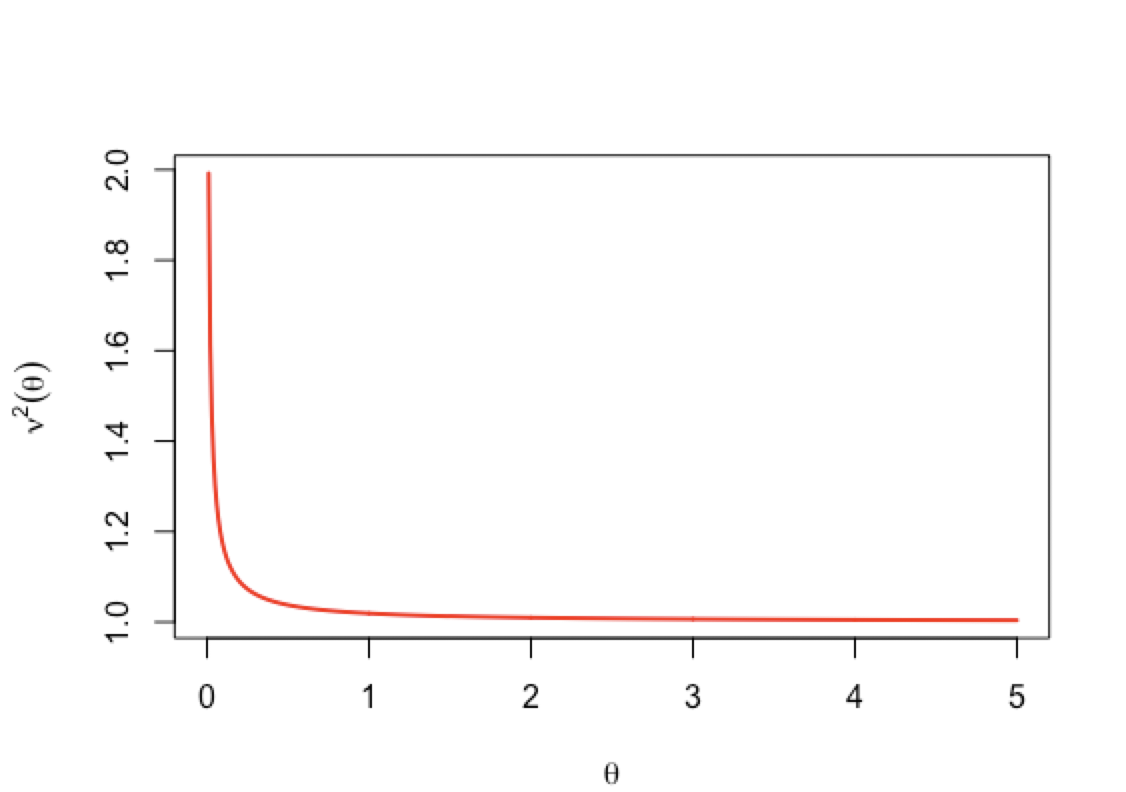}
\caption{Worst-case volatility ratio $\nu^2(\theta)$ as a function of $\theta$. }
\end{figure}
The results of proposition \ref{s2} can be seen in figure \ref{figs2}. 
This result shows that the investor's implied beliefs converge to the benchmark model as their model confidence becomes infinite. Together with result $(iii)$ of the previous proposition, this formally establishes the subjective expected utility model without ambiguity aversion as a limit of the problem with ambiguity aversion. 

\subsubsection{Comparison with ``unconstrained'' or ``multiplier'' preferences}
It is natural to compare the solution to the portfolio choice problem in the previous section to the corresponding ``unconstrained'' problem where we ignore the mean restriction $\mathbb{E}_P[M (\widetilde{R} - \mu)] = 0$. This corresponds to the portfolio choice of an ambiguity-averse investor with quadratic utility and ``multiplier'' preferences. 

Section \ref{appen:multiplier} of the appendix describes the solution to the portfolio choice problem in the previous section ignoring the mean restriction. The unrestricted problem has similar features to the restricted problem. In particular, under the implied worst-case $M$, the return on wealth is still normally distributed. However, both the mean and variance are distorted and will depend on the penalty parameter $\theta$. For regions of the penalty parameter $\theta$ where a solution exists, the distorted mean $\mu_u(\theta)$ is increasing in $\theta$, while the distorted variance $\sigma_u^2(\theta)$ is decreasing in $\theta$. The optimal portfolio weight on the risky asset is
\[
\phi_u(\theta) = \frac{1}{\mu_u(\theta)^2 + \sigma_u^2(\theta)} \mu_u(\theta) (b - W_0 R_f). 
\]
Results analogous to proposition \ref{s1} hold for $\phi_u(\theta)$, albeit for slightly different reasons. Under moment-constrained, increasing $\theta$ decreases the implied variance of the return under the worst-case distribution. In the unconstrained problem, both the mean and variance change with $\theta$. 

\section{Dynamic preferences}
Next, I present a dynamic extension to the static preferences defined by equation \eqref{pref1}. Consider a $J+1$ period discrete-time setting. Denote the time between periods by $\Delta$ so time $t$ is discrete and satisfies $t \in \{0,...,J\Delta\}$. I assume that the decision-maker's utility in period $t+1$ depends only on the time-$t$ and $t+1$ realizations of a stochastic process $\{X_t\}_{t \in \{0,...,J\Delta \} }$. For simplicity, I abstract from modelling how the decision-maker's choices affect $X$ and simply define utility taking $X$ to be an exogenous, time homogeneous Markov process with Gaussian increments under the benchmark probability measure $P$, i.e.
\begin{equation}
X_{t+\Delta} - X_t = \mu(X_t) \Delta + \sigma(X_t) \epsilon_{t+\Delta}
\end{equation}
where $\epsilon_{j\Delta} \overset{iid}{\sim} \text{Normal}(0,\Delta)$ under $P$ for all $j \in \{1,...,J\}$. As in the previous section, we consider absolutely continuous changes of measure parameterized by positive random variables $M$ with unit expectation. For each $M$, define
\[
M_t = \mathbb{E}_t \left[ M \right] 
\]
so that $M_t$ is a positive martingale relative to the filtration generated by the process $\{X_t\}_{t=0}^T$. Additionally, define
\[
M_{t,t+\Delta} = \frac{M_{t+\Delta}}{M_t}. 
\]
Note that $M_{t,t+\Delta}$ is a positive random variable with time-$t$ conditional expectation 1. Observe that $M_{t,t+\Delta}$ can be interpreted as a conditional change-of-measure or conditional likelihood ratio. \\

While the decision maker does not have full confidence in $P$, she is certain about specific conditional moments of the data-generating process. Analogous to the moment restriction in the static model, the decision maker will only consider models generated by martingale distortions $M$ which satisfy the moment restriction
\begin{equation}
\label{conditionalmoment}
\mathbb{E}_t \left[ M_{t,t+\Delta} \left( X_{t+\Delta} - X_t - \mu(X_t) \Delta  \right) \right] = 0, \ \forall t. 
\end{equation}
Equation \eqref{conditionalmoment} captures the decision-maker's certainty about the expected one-period change in the process $X_t$. Thus, the decision-maker's uncertainty is limited to uncertainty about higher moments of the distribution, such as second moments or variances. Note that equation \eqref{conditionalmoment} is satisfied by the constant random variable $M = 1$. 

Write the time-$0$ utility of the decision-maker as
\begin{equation}
\label{dyn1}
V_0(X_0;P,\theta) = \inf_{ M \geq 0, \mathbb{E}[M] = 1} \mathbb{E}_0 \left[ M_{(j+1)\Delta} \sum_{j=1}^J e^{-\rho j \Delta} u(X_{j\Delta},X_{(j+1)\Delta} )\Delta  + c(M) \right] 
%V_0(X_0; P,\theta) = \inf_{M \geq 0, \mathbb{E}[M] = 1} \mathbb{E}_0 \left[ M \sum_{t=1}^T \beta^t u(X_t,X_{t-1})   + c(M) \right] 
\end{equation}
where
\[
c(M) = \begin{cases}
\theta(\Delta) M_{(j+1)\Delta} \sum_{j=1}^J e^{-\rho j \Delta}  \log (M_{j\Delta,(j+1)\Delta}) & \text{ if } \eqref{conditionalmoment} \text{ holds for all } j \in \{1,...,J\} \\
\infty & \text{ otherwise }
\end{cases}
\]
It can easily be shown that the ambiguity index $c(\cdot )$ is convex in $M$. Therefore, this preference specification fits into the very general framework of dynamic variational preferences proposed and axiomatized by \cite{maccheroni2006dynamic}. The particular form of the ambiguity index considered here is the same as the \emph{discounted relative entropy} penalty used in the multiplier preferences of \cite{hansensargent:2001} and followup papers but with an added sequence of conditional moment restrictions on $M$. 

\cite{CHH2020} study a mathematically similar problem that arises when an econometrician is interested in bounded implied subjective expectations of economic agents subject to a vector of conditional moment conditions, meant to convey partial information about asset prices or survey data on subjective expectations. In particular, their problem can be thought of as an Ergodic control problem that arises in the limit as $\rho \to \infty$.

\subsection{Continuous-time limit}
Next, I give a heuristic derivation of a continuous-time limit of \eqref{dyn1}. The corresponding limit will be used as preferences in subsequent sections where I consider models of repeated moral hazard. In these settings, the use of continuous-time diffusion models greatly improves tractability of the optimal contracting problems. 

We will take the limit of the discrete-time problem in the previous section as $\Delta \to 0$. Note that the discrete-time process, $X_t$ will converge in law to a continuous-time diffusion process with evolution equation
\[
dX_t = \mu(X_t) dt + \sigma(X_t) dZ_t
\]
where $Z_t$ is a standard Brownian motion. 

Let $V_t(X_t)$ denote the time-$t$ continuation value function of the decision-maker in the discrete-time problem. Note that we have the following Bellman-type equation for $V_t$,
\begin{align*}
V_t(X_t) = &\inf_{m \geq 0, \mathbb{E}_t[m] = 1} \mathbb{E}_t \left[ m u(X_{t+\Delta},X_t) + \theta(\Delta) m \log m + e^{-\rho \Delta} V_{t+\Delta}(X_{t+\Delta}) \right] \\
& \text{ subject to  } \mathbb{E}_t[ m \left( X_{t+\Delta} - X_t - \mu(X_t) \Delta \right)  ] = 0
\end{align*}
where the choice variable $m$ is the conditional one-period likelihood ratio chosen by the adversarial nature. 
Under mild regularity conditions, the functions $u(\cdot,\cdot)$ and $V_t(\cdot)$ will be twice continuously differentiable. Therefore, we can approximate them as local quadratic functions in the increment $X_{t+\Delta} - X_t$. Under this approximation, we then have that the worst-case one-period likelihood ratio $m$ will have the form
\[
m^* = m(\epsilon_{t+\Delta},X_t) = \frac{\exp \left(  - \frac{1}{2}(\epsilon_{t+\Delta})^2 \omega(X_t) \right) }{\mathbb{E}_t \left[ \exp \left(  - \frac{1}{2} (\epsilon_{t+\Delta})^2 \omega(X_t) \right)   \right]  }
\]
where $\omega(\cdot )$ is a function of $X_t$ that depends implicitly on $\Delta, \theta, \rho$, the curvature of $u(\cdot, X_t)$, and  $V_{t+\Delta}(\cdot)$. As before, this is easily obtained from convex duality results. We see that the implied $m^*$ will change the variance of the Gaussian shock $\epsilon_{t,t+\Delta}$ from 1 to some unknown function $\nu^2(X_t)$. We can then equivalently think of the choice of likelihood ratio $m$ as being a choice of change-of-volatility $\nu$. 

As a function of $\nu$, the one-period relative entropy of a volatility distortion is given by
\[
\mathbb{E}[ m(\nu) \log m(\nu) ] = \frac{1}{2} \left\{ \nu^2 - 1 - \log \nu^2  \right\}
\]
Taking the limit as $\Delta \to 0$, and the number of periods $J \to \infty$ so that $J \Delta \to T$ and letting $\theta(\Delta) = \theta \Delta$, we obtain that time-0 lifetime utility will be given by
\begin{equation}
\label{cts1}
V_0(X_0;P,\theta) = \inf_{ \{\nu_t\}_{t=0}^T} \mathbb{E}_0 \left[ \int_0^T e^{-\rho t} \left( u(X_t)  + \frac{\theta}{2} \left\{ \nu_t^2 - 1 - \log \nu_t^2 \right\} \right) dt    \right] 
\end{equation}
where the infimum is subject to the constraint
\[
dX_t = \mu(X_t) dt + \sigma(X_t) \nu_t dZ_t.
\]
I make two important observations. First, the infimum problem in equation \eqref{cts1} is a different problem from the infimization in the discrete-time problem in equation \eqref{dyn1}. In discrete-time, preferences were defined as an infimum over probability distributions, whereas now we are representing preferences as an infimum over a controlled process. Thus the continuous-time limit here is best thought of as an equivalent problem that produces the same value function.\footnote{Similar considerations arise for continuous-time multiplier preferences.} Second, the linear scaling $\theta(\Delta) = \theta \Delta$ is important for the limit. The standard continuous-time limit for multiplier preferences would let $\theta(\Delta)$ be constant as $\Delta \to 0$. This imposes absolute continuity of measures in the continuous-time limit, which by Girsanov's theorem restricts all probabilistic distortions to conditional drift distortions. By contrast, the limit I consider allows violations of absolute continuity in the continuous-time setup. If it weren't for the conditional moment restrictions, this would allow the adversarial nature to choose arbitrarily large drift distortions at zero cost. I refer to any process $\{\nu_t\}_{t=0}^T$ that attains the infimum in \eqref{cts1} as a \emph{worst-case} change-of-volatility process. 

Additionally, observe that for the function $V_t(X_t;P,\theta)$ we will have the following PDE,
\[
0 = \min_{\nu} \ u(X) + \frac{\theta}{2} \left\{ \nu^2 - 1 - \log \nu^2 \right\} + \frac{\partial V}{\partial t} - \rho V + \frac{\partial V}{\partial X} \mu(X) + \frac{1}{2} \frac{\partial^2 V}{\partial X^2} \sigma^2(X) \nu^2. 
\]
For simplicity, I will consider the infinite-horizon limit as $T \to \infty$ so that the problem will become stationary and therefore $\frac{\partial V}{\partial t} = 0$. 

%Note that the effect of ambiguity aversion does not vanish in the continuous-time limit here as $\Delta \to 0$. This is distinct from \cite{skiadas2013smooth} who argued that such effects should vanish in the continuous-time limit of distinct but related smooth ambiguity models. This discrepancy effectively arises over the treatment of the parameter $\theta(\Delta)$ in the limit as $\Delta \to 0$. See \cite{hansen2018aversion} for further discussion of these issues. 

\subsubsection{Observability of volatility and calibration of $\theta$}

The penalty parameter $\theta$ captures the degree of confidence that the economic actor has in their benchmark model of volatility, with $\theta = \infty$ corresponding to complete model confidence, i.e. expected utility. 

It is well-known that continuous-time diffusion models imply that volatility is directly observable via the quadratic variation. One might conclude that this implies that the only reasonable value for $\theta$ is infinity. I give several reasons why this is not the case in many domains of interest:
\begin{enumerate}[(i)]
\item Across empirical domains, despite numerous proposed models of stochastic volatility, there is generally no accepted consensus on the ``correct'' parametric model for stochastic volatility. It seems sensible therefore that economic actors would take any parametric model of volatility as an approximation. 
\item Many economically important quantities (such as accounting variables, macroeconomic growth rates, inflation, values of illiquid assets) are not observed at high frequencies. Diffusion models applied to these settings are approximations with desireable tractability features, and thus the implication of observable volatility should not be taken literally. 
\item Even in asset pricing settings where high-frequency data is readily available, the literature on high-frequency econometrics (see for instance \cite{zhang2005tale}, \cite{bandi2006separating}, \cite{hansen2006realized}, and \cite{bandi2008microstructure}) finds significant statistical evidence of ``microstructure noise'', i.e. that the latent price process (assumed to be a semimartingale) is contaminated by weakly-dependent measurement error due to liquidity or trading frictions. This implies that the integrated quadratic variation is different from the integral of the true latent volatility process, and that volatility is not directly observable. 
\end{enumerate}
If we accept that economic actors do not observe volatility directly for reasons (ii) or (iii), then we may follow the robust control literature and apply detection error probabilities to calibrate ``reasonable'' values for $\theta$ by either adopting a fixed $\Delta > 0$ convention or appending a measurement error process to a continuous-time volatility specification (see for instance \cite{anderson2003quartet} and \cite{hansen2002robust} for calibrations based on detection error probabilities). Alternatively, we may apply the subjective reasoning of \cite{good1952rational} and consider a value of $\theta$ as ``reasonable'' if the implied worst-case model is subjectively reasonable.

\subsection{Comparison with $G$-expectations}
An alternate approach to modelling ambiguous volatility in continuous-time was introduced by \cite{peng2007g}. This approach, known as $G$-expectations, can be thought of as a continuous-time counterpart to the max-min expected utility of \cite{Gilboa_Schmeidler} applied to an unknown volatility parameter. The implications of this approach are explored by \cite{epstein2013ambiguous} in an asset pricing setting. As will be demonstrated, the $G$-expectations approach is closely related to the approach described in the previous section. 

Consider preferences defined by the following generalization of the continuous-time preferences defined by \eqref{cts1}
\[
U(X_0) = \inf_{ \{\nu_t\}_{t=0}^\infty} \mathbb{E}_0 \left[ \int_0^\infty e^{-\rho t} \left( u(X_t)  + \xi(\nu_t) \right) dt    \right] 
\]
where $\xi(\cdot)$ is a convex function of $\nu$ which I refer to as the penalty function. Note that these preferences imply the following PDE for $U$,
\[
0 = \min_{\nu} \ u(X) + \xi(\nu) - \rho U + \frac{\partial U}{\partial X} \mu(X) + \frac{1}{2} \frac{\partial^2 U}{\partial X^2} \sigma^2(X) \nu^2. 
\]
Of course, the preferences in \eqref{cts1} can be seen to be a special case by taking $\xi (\nu ) = \frac{\theta}{2} \left\{ \nu^2 - 1 - \log ( \nu^2 ) \right\}$. $G$-expectations can also be thought of as a special case, but now by taking $\xi(\nu)$ to be a convex indicator function. In the case where the benchmark volatility $\sigma(X) = \sigma$ is constant, this corresponds to
\[
\xi(\nu) = \begin{cases}
0 &\text{ if } \nu \in \left[ \underline{\sigma}/\sigma, \overline{\sigma}/\sigma \right] \\
\infty &\text{ otherwise. }
\end{cases}
\]
Note that here the convex indicator function restricts the instantaneous volatility $\sigma \nu_t$ under the worst-case model to be in the interval $[\underline{\sigma}, \overline{\sigma}]$. The differences between these two approaches are analogous to the differences between max-min expected utility and variational or multiplier formulations of ambiguity aversion. 

Obviously these two approaches are mathematically similar. Nonetheless there will be differences in implications of the two models, some of which will be explored in the subsequent application. In particular, while the two models will have qualitatively similar implications for the optimal contract, they will have markedly different implications for the worst-case volatility model and asset prices. For the optimal contracting problem I consider in the following section, the $G$-expectations model imply a constant worst-case volatility of $\sigma \nu_t = \overline{\sigma}$ whereas the relative entropy model will imply a worst-case volatility process that is state-dependent and higher in states that are worse for the decision-maker.

\begin{remark}
As \cite{peng2007g} and \cite{nutz2013random} demonstrate, ambiguity about volatility in diffusion environments can be represented via a mathematically convenient control theory implementation. The implied value function and HJB equation are the same as one in which the unknown volatility is treated as a controlled process. This paper does not formally extend their equivalence results, and sidesteps this by simply studying the equivalent control problem. Rigorous development of nonlinear expectation theory extending the equivalence results of  \cite{peng2007g} and \cite{nutz2013random} to cover the applications considered here is well beyond the scope of this paper. 
\end{remark}

%%%% SCALING AND ABSOLUTE CONTINUITY
% Anderson, Hansen, and Sargent
% Miao and Rivera
% Skiadas scaling eliminates all effects of ambiguity aversion

\section{Application: Optimal security design under repeated moral hazard}
To illustrate the effect of the continuous-time preferences derived in the previous section, I apply them to a model of optimal contracting under repeated moral hazard based on papers by \cite{demarzo2006optimal} and \cite{biais2007dynamic}.

\subsection{Setup}
I first describe a benchmark model without ambiguity based on \cite{demarzo2006optimal} and \cite{biais2007dynamic}. At each instant $t$, agent chooses an effort level $a_t \in [0,1]$. Given an effort choice, the cumulative cash-flow process $\{Y_t\}$ obeys the law of motion
\begin{equation}
\label{be}
dY_t = \mu a_t dt + \sigma dZ_t
\end{equation}
where $\mu,\sigma > 0$, and $Z_t$ a standard Brownian motion. 

The agent can derive private benefits $\lambda \mu (1-a_t)$ from the action $a_t$ where $\lambda \in (0,1)$. Due to linearity, it is without loss of generality to take $a_t \in \{0,1\}$. At any time $t \geq 0$ the project can be liquidated, producing a liquidation value of $L$. The principal and the agent are both assumed to be risk neutral. The principal discounts cash flows at a rate $r > 0$ while the agent discounts cash flows at a rate $\gamma > r$.\footnote{This assumption means that the agent is impatient relative to the principal, and avoids degeneracy.}

In selecting an optimal contract, the principal chooses a cumulative compensation process $C$ for the agent, a liquidation stopping time $\tau$, and a  suggested effort process $a$ for the agent. The benchmark model optimal contracting problem is given as follows. 
\begin{problem}[benchmark model]
\begin{equation}
\max_{(C,\tau,a)} \mathbb{E}^{P^a} \left[\int_0^\tau e^{-rs} (dY_s - dC_s ) + e^{-r\tau} L   \right] 
\end{equation}
subject to
%Need to re-write as align with labels... (figure this out at some point)

\begin{align}
\mathbb{E}^{P^a} \left[ \int_0^\tau e^{-\gamma s} (dC_s + \lambda \mu (1-a_s) ds)  \right] 
&\geq \mathbb{E}^{P^{\widehat{a}}} \left[ \int_0^\tau e^{-\gamma s} (dC_s + \lambda \mu (1-\widehat{a}_s) ds)  \right] \label{IC} \\
\mathbb{E}^{P^a} \left[ \int_0^\tau e^{-\gamma s} (dC_s + \lambda \mu (1-a_s) ds)  \right] &= W_0. \label{IR}
\end{align}

%\begin{equation}\label{IC}
%\mathbb{E}^{P^a} \left[ \int_0^\tau e^{-\gamma s} (dC_s + \lambda \mu (1-a_s) ds)  \right] 
%\geq \mathbb{E}^{P^{\widehat{a}}} \left[ \int_0^\tau e^{-\gamma s} (dC_s + \lambda \mu (1-\widehat{a}_s) ds)  \right] 
%\end{equation}
%\begin{equation}\label{IR}
%\mathbb{E}^{P^a} \left[ \int_0^\tau e^{-\gamma s} (dC_s + \lambda \mu (1-a_s) ds)  \right] = W_0
%\end{equation}
\end{problem}

\subsection{Optimal Contract}
Assume for simplicity that only the principal is ambiguity-averse. This will turn out to be without loss of generality. The principal takes the evolution equation \eqref{be} as an approximate benchmark model, but allows for alternate models where the cumulative cash flow process evolves as
\begin{equation}
\label{vol}
dY_t = \mu a_t dt + \sigma \nu_t dZ_t. 
\end{equation}
To avoid a degenerate effect of ambiguous volatility, it is necessary to assume that realized volatility is not directly contractible. Otherwise it would be possible for the agent to fully insure the principal's uncertainty about volatility without any subjective welfare loss. I formalize this with the following restriction.  

\begin{Restriction}
\label{restrict:mrt}
Under any feasible contract $(C,\tau,a)$, the process 
\begin{equation}
M_t = \mathbb{E}_t^{P,a} \left[ \int_0^\tau e^{-\gamma s} ( dC_s + \lambda \mu (1-a_s) ds  \right] 
\end{equation}
admits the martingale representation
\begin{equation}
\label{eq:mrt}
M_t = M_0 + \int_0^t \phi_s (dY_s - \mu a_s ds ). 
\end{equation}
under $P$. 
\end{Restriction}
If there were no uncertainty about volatility, then equation \eqref{eq:mrt} in restriction \ref{restrict:mrt} would simply follow from the martingale representation theorem. However, since the principal and agent have potentially different beliefs about volatility, the principal and agent could derive subjective welfare improvements from writing contracts in which the agent's compensation is contingent on the realized volatility. This is disallowed by restriction \ref{restrict:mrt}. 

%Maybe first characterize IC condition as in DeMarzo Sannikov

The optimal contracting problem is given by:
\begin{problem}\label{contract}
\begin{equation}
\sup_{(C,\tau,a)} \inf_{\nu} \mathbb{E}^{\nu} \left[  \int_0^\tau e^{-rt} (dY_t - dC_t) + e^{-r\tau} L  \right] +  \mathbb{E}^{\nu} \left[  \int_0^\tau e^{-rt} \xi (\nu_t) dt   \right] 
\end{equation}
subject to equations \eqref{IC}, \eqref{IR}, \eqref{vol}, and restriction \ref{restrict:mrt}. 
\end{problem}

Problem \ref{contract} can be thought of as a two-player, zero-sum stochastic differential game\footnote{See \cite{fleming1989existence} for further discussion.} between the principal and an adversarial nature. Nature chooses the time-varying change of volatility process $\nu_t$ to minimize the welfare of the agent, but choosing $\nu_t$ different from one has cost proportional to the instantaneous relative entropy. 
%Next, I heuristically derive the Hamilton-Jacobi-Bellman-Isaacs (HJBI) equation for optimality. 

%\textcolor{red}{REMOVE THIS STUFF:
%Let $\phi_t$ be the sensitivity to $\nu_t \sigma d B_t^\nu$ i.e. the cash-flow shock under the probability measure $Q^\nu$. By the martingale representation theorem, $W_t$ satisfies
%\begin{equation}
%\label{mrt}
%d W_t = \gamma W_t dt - dC_t - \lambda \mu_t (1-a_t) dt + \phi_t  \sigma \nu_t d B_t^\nu
%\end{equation}
%Note that equation \eqref{mrt} implicitly assumes that the worst-case volatility $\sigma \nu_t$ is not directly contractable. 
%}

%%%% PUT SOMETHING ABOUT CONTINUATION PAYOFF OF AGENT AS STATE VARIABLE HERE!!!!!

Let $W_t$ denote the time-$t$ continuation payoff of the agent. It follows from restriction \ref{restrict:mrt} that
\begin{equation}
dW_t = \gamma W_t dt - dC_t - \lambda \mu_t (1-a_t) dt + \phi_t (dY_t - \lambda \mu (1-a_t) dt )
\end{equation}
Observe that in view of equation \eqref{vol}, the principal and the agent perceive the evolution of $W_t$ differently. The principal perceives it as
\begin{equation}
dW_t = \gamma W_t dt - dC_t - \lambda \mu_t (1-a_t) dt + \phi_t \sigma \nu_t dZ_t
\end{equation}
whereas the agent perceives it as
\begin{equation}
dW_t = \gamma W_t dt - dC_t - \lambda \mu_t (1-a_t) dt + \phi_t \sigma dZ_t. 
\end{equation}

\subsubsection{First-Best Contract}
The first-best contract is the same as the first-best contract with no ambiguity aversion in \cite{demarzo2006optimal}. This is intuitively obvious since the first-best value function is linear, hence there are no volatility effects. 

\begin{equation}
r F(W) = \sup_{c \geq 0, \phi } \inf_{\nu} \mu - c +\psi(\nu) + \left( \gamma W  - c \right)F'(W)  + \frac{1}{2}   \phi^2 \nu^2 \sigma^2 F''(W). 
\end{equation}
It is easy to verify that at the optimum, we have $\phi=0$ and therefore the principal's value function under the (stationary) first-best contract is
\[
F(W) = \frac{\mu}{r} - \frac{\gamma}{r} W,
\]
which can be implemented by the principal paying the agent a constant wage of $c = \gamma W$. Of course, this can be improved if we allow time-0 lump sum transfers in which case the principal can simply give a one-time transfer of $W$ to the agent which gives
\[
F(W) - W = \frac{\mu}{r}. 
\]
Thus with no moral hazard, volatility ambiguity produces no reduction in welfare.

\subsubsection{Optimal Contract with Moral Hazard}
%Maybe do formal result here!!!!
It is a simple extension of lemma 3 of \cite{demarzo2006optimal} to show that for any change-of-volatility process $\nu_t$, the agent's incentive compatibility constraint can be written as
\begin{equation}
\phi_t \geq \lambda
\end{equation}
The HJBI equation for the optimal contract with agency is given by
\begin{equation}
r F(W) = \sup_{c \geq 0, \phi \geq  \lambda } \inf_{\nu} \mu - c + \frac{\theta}{2} \left\{ \nu^2 - 1 - \log (\nu^2)  \right\} + \left( \gamma W  - c \right)F'(W)  + \frac{1}{2}   \phi^2 \nu^2 \sigma^2 F''(W). 
\end{equation}
A simple calculation shows that the worst-case change of volatility $\nu$ is given by
\begin{equation}
\nu^2 = \frac{\theta}{\theta + \phi^2 \sigma^2 F''(W)}. 
\end{equation}

Plugging in the our expression for $\nu^2$, the HJBI reduces to the following nonlinear HJB equation
\begin{equation}\label{HJB}
r F(W) = \sup_{c \geq 0, \phi \geq \lambda} \mu - c - \frac{\theta}{2} \log(\theta)    + \left( \gamma W - c \right) F'(W)  + \frac{\theta}{2}  \log ( \theta + \phi^2 \sigma^2 F''(W)). 
\end{equation}

Consider the region $[0,\overline{W})$ for which $F'(W) > -1$ so that $c=0$ is optimal. 
Rearranging \eqref{HJB} gives
\[
\sup_{\phi \geq \lambda} \frac{\theta}{2} \log \left( 1 + \frac{\phi^2 \sigma^2}{\theta} F''(W)   \right) = r F(W) - \mu - \frac{\theta}{2} - \gamma W F'(W)
\]
Now I apply $rF(W) - \mu \leq \gamma W$ which comes from the second-best value function being less than or equal to the first-best value function without lump-sum transfers, and $F'(W) > -1$ to obtain
\[
\sup_{\phi \geq \lambda} \frac{\theta}{2} \log \left( 1 + \frac{\phi^2 \sigma^2}{\theta} F''(W)   \right) < 0
\]
which is a contradiction unless $F''(W) < 0$. This shows that $F$ is strictly concave on $[0,\overline{W}]$. Intuitively this holds because unnecessarily exposing the agent to cash flow shocks is costly to the principal, since it increases the probability of inefficient liquidation.

Thus we have shown the following. On the interval $[0,\overline{W}]$, the principal's value function satisfies the ODE
\[
r F(W) = \mu  + \gamma W F'(W) + \frac{\theta}{2} \log\left( 1 + \frac{\lambda^2 \sigma^2}{\theta} F''(W)\right). 
\]
$F$ is strictly concave so the worst-case change of volatility given by
\begin{equation}\label{nu}
\nu^*(W)^2 = \frac{\theta}{\theta + \lambda^2 \sigma^2 F''(W)}
\end{equation}
is strictly greater than 1. Additionally, the strict concavity of the value function implies that the incentive constraint always binds, i.e.
\[
\phi^*(W) = \lambda. 
\]
While this is the same as \cite{demarzo2006optimal}, it stands in contrast to \cite{miao2016robust}. 

The next proposition characterizes the optimal contract under the assumption that high effort is always optimal. The optimal contract with partial shirking can be described using methods similar to \cite{zhu2013optimal}, but such a characterization is beyond the scope of this paper. 

\begin{proposition}
\label{main}
Assume that $L < \frac{\mu}{r}$ and that implementing high effort is optimal. Assume further that there exists a unique twice differentiable solution $F$ to the ODE
\begin{equation}\label{ode}
r F(W) = \mu + \gamma W F'(W) + \frac{\theta}{2} \log\left( 1 + \frac{\lambda^2 \sigma^2}{\theta} F''(W)\right)
\end{equation}
with boundary conditions
\[
F(0) = L, , \ F'(\overline{W}) = -1
\]
and $F''(W) < 0$ for all $W \in [0,\overline{W})$ where $\overline{W}$ is defined by $F''(\overline{W}) =0$. % This implies F''(\overline{W}) = 0
Then:
\begin{enumerate}[(i)]
\item When $W \in [0,\overline{W}]$, $F(W)$ is the principal's value function for problem \ref{contract}, the optimal cash flow sensitivity is $\phi^*(W) = \lambda$ and the worst case change of volatility $\nu^*(W)$ is given by \eqref{nu}. The contract delivers value $W$ to the agent whose continuation value $W_t$ evolves according to
\[
dW_t = \gamma W_t dt - dC_t^* + \phi^*(W_t) \sigma \nu^*(W_t)\ dZ_t
\]
where $dC_t^*$ is 0 in $[0,\overline{W})$ and causes $W_t$ to reflect at $\overline{W}$. The contract terminates at time $\tau = \inf\{t \geq 0 : W_t = 0\}$ when the project is liquidated. 
\item When $W > \overline{W}$, the principal's value function is $F(W) = F(\overline{W}) - (W - \overline{W})$. The principal immediately pays $W - \overline{W}$ to the agent and contracting continues with the agent's new initial value $\overline{W}$. 
\end{enumerate}
\end{proposition}

Observe that as the degree of model confidence $\theta \to \infty$, the quantity $\frac{\theta}{2} \log \left( 1 + \frac{\lambda^2 \sigma^2}{\theta} F''(W)  \right)$ converges to $\frac{1}{2} \lambda^2 \sigma^2 F''(W)$ for any value of $F''(W)$. Hence \eqref{ode} converges to the ordinary differential equation of \cite{demarzo2006optimal}, i.e. the benchmark model with no ambiguity aversion. 

\begin{proposition}
Let $F(\cdot )$, $\overline{W}$ be defined as in proposition \ref{main}. Then high effort is optimal if and only if
\[
\min_{W \in [0,\overline{W}]} rF(W) - F'(W) (\gamma W - \lambda \mu) \geq 0. 
\]
\end{proposition}

%A simple argument now shows that implementing high effort is optimal if and only if

The argument is simple, and is consistent with proposition 8 of \cite{demarzo2006optimal}. Next, I show how the optimal contract changes with the level of ambiguity aversion.

\begin{proposition}\label{welfare}
For any promised wealth level to the agent, the principal's value function $F(W)$ strictly increases in $\theta$. 
\end{proposition}

\begin{figure}[h!]
\centering
\includegraphics[scale=0.5]{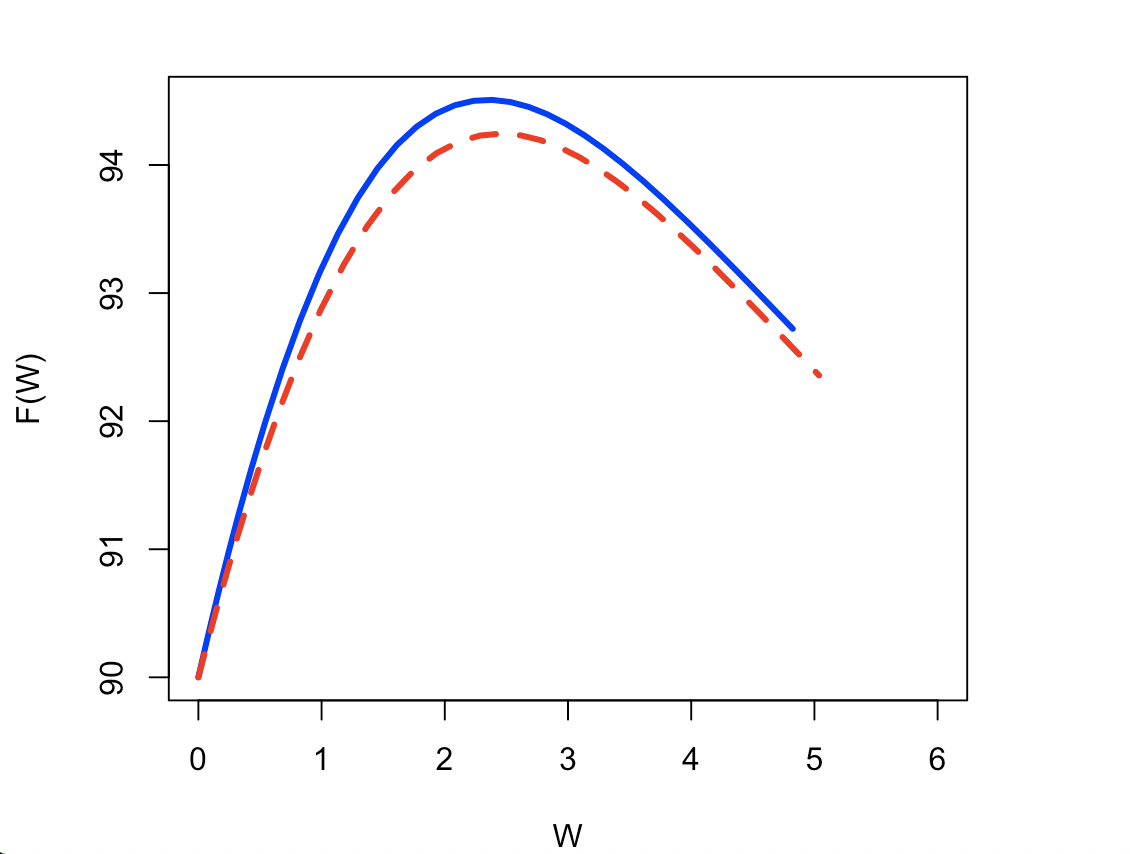}
\caption{Value functions $F(W)$ for contracting problem. Value function with no ambiguity ($\theta = \infty$) shown in blue. Value function with $\theta = 5$ shown in dashed red. Parameter values are $\mu = 10, r = 0.1, \gamma = 0.15, \lambda = 0.2, \sigma = 5, L = 90$. Observe that value function with no ambiguity is strictly higher than value function with $\theta = 5$, consistent with proposition \ref{welfare}. }
\end{figure}

\begin{figure}[h!]
\centering
\includegraphics[scale=0.5]{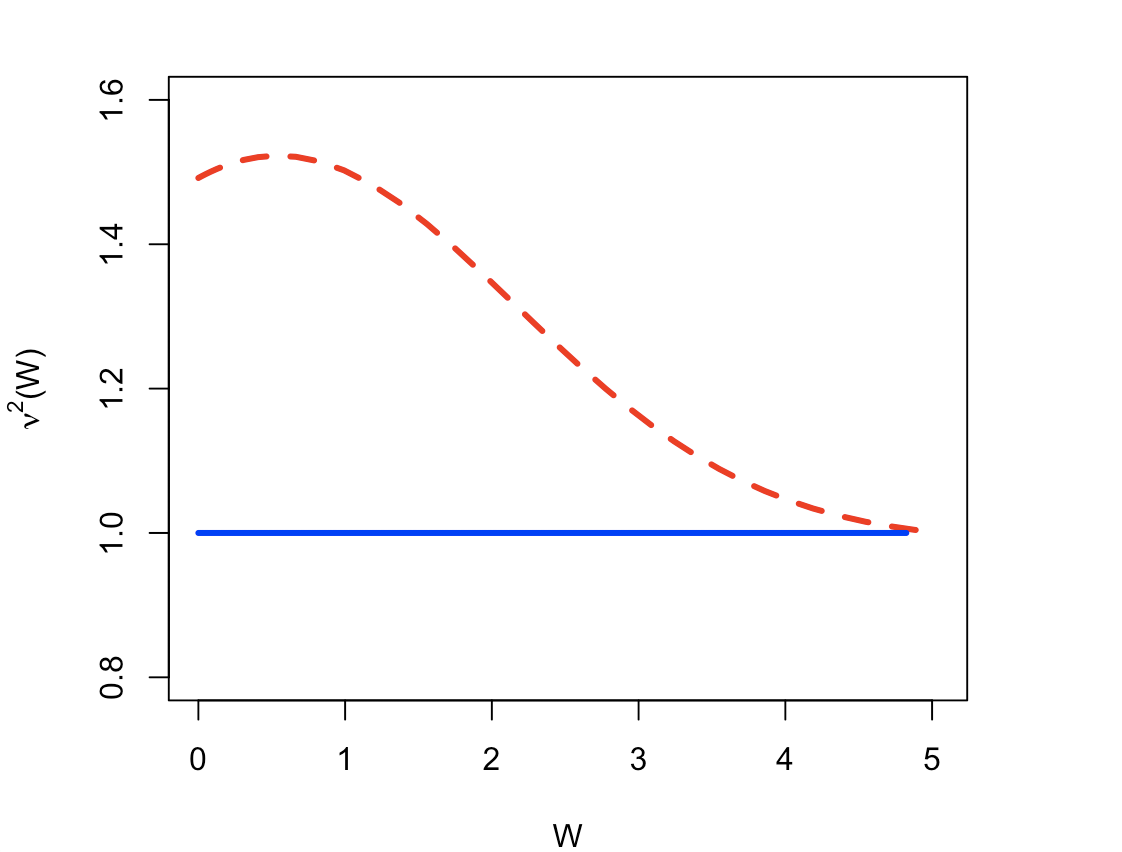}
\caption{Worst case change-of-variance $\nu^2(W)$ for $\theta = 5$ shown in red. $\nu^2 = 1$, i.e. change-of-variance when $\theta = \infty$ shown in blue. Parameter values are $\mu = 10, r = 0.1, \gamma = 0.15, \lambda = 0.2, \sigma = 5, L = 90$. }
\end{figure}

\begin{figure}[h!]
\centering
\includegraphics[scale=0.5]{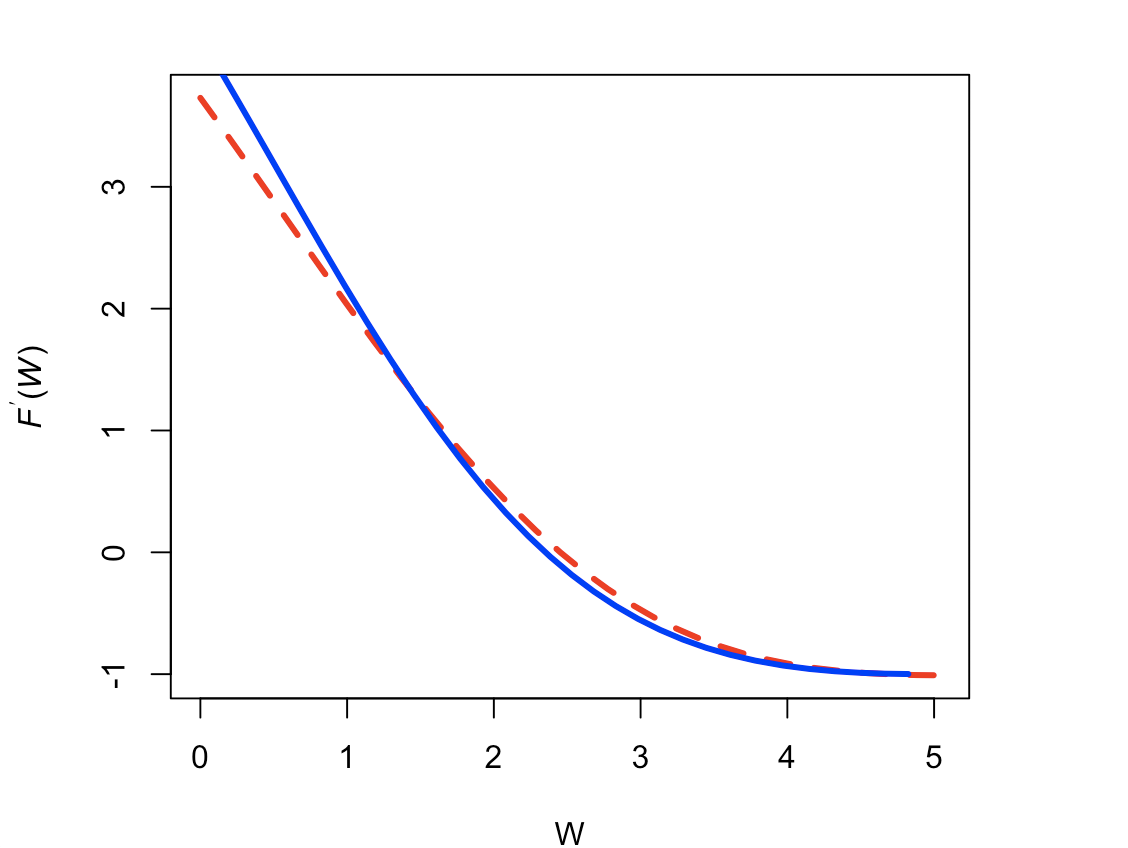}
\caption{Value function derivative $F'(W)$ for $\theta = 5$ shown in dashed red line. Value function derivative with no ambiguity ($\theta = \infty$) shown in blue. Parameter values are $\mu = 10, r = 0.1, \gamma = 0.15, \lambda = 0.2, \sigma = 5, L = 90$. Points for which $F'(W) = -1$ correspond to upper boundary $\overline{W}$.}
\end{figure}

Proposition \ref{welfare} confirms the obvious intuition that the principal's value function is increasing in $\theta$ i.e. decreasing in the level of ambiguity aversion. This is illustrated in figure 1. While this result is unsurprising, it is nonetheless useful in establishing subsequent comparative static results for the optimal contract. The following proposition shows how the payoff boundary $\overline{W}$ changes with $\theta$. 

\begin{proposition}\label{payoffboundary}
The payoff boundary $\overline{W}$ is strictly decreasing in $\theta$.  
\end{proposition}

\begin{figure}[h!]
\centering
\includegraphics[scale=0.5]{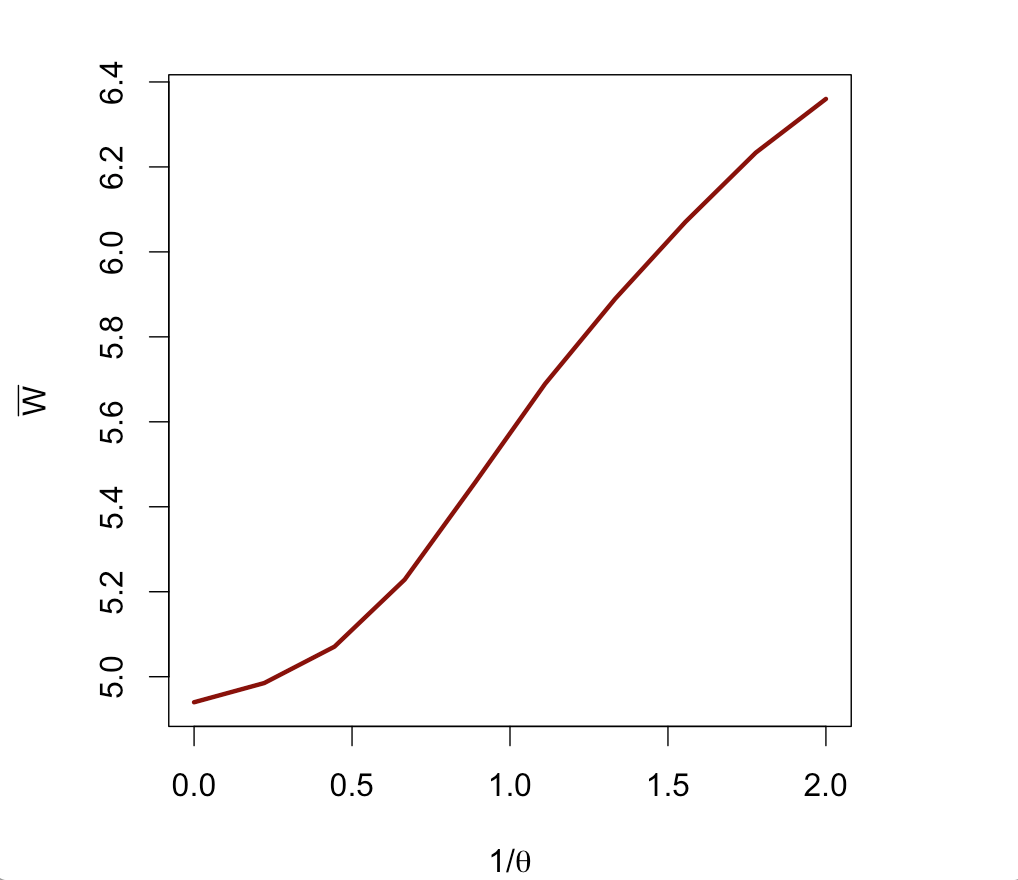}
\caption{Upper payoff boundary $\overline{W}$ as a function of ambiguity aversion $1/\theta$. Parameter values are $\mu = 10, r = 0.1,\gamma = 0.15,\lambda = 0.2, \sigma = 5, L = 90$.  }
\label{figure:wbar}
\end{figure}

Thus, higher levels of ambiguity aversion leads to a higher payoff boundary for the agent.  This result is illustrated in figure \ref{figure:wbar}. 

\FloatBarrier

\subsubsection{What happens if the agent is ambiguity-averse?}

Up until this point, I have assumed that only the principal was ambiguity averse. It is natural to ask what happens if the agent is ambiguity averse as well. As it turns out, so long as the agent has the same form of variational ambiguity with a strictly convex penalty function, the agent's ambiguity aversion will not affect the optimal contract. 

\begin{proposition}\label{agent}
Assume that the agent is ambiguity averse. Then the contract described in proposition \ref{main} remains optimal. Moreover, the agent's implied worst-case belief is $\nu(W) = 1$. 
\end{proposition}

Even when the agent is ambiguity averse, the optimal contract is unaffected and they form expectations as if they fully trust that volatility is constant at level $\sigma$. 

\subsubsection{Bellman-Isaacs condition}

The optimal contract characterized by proposition \ref{main} is the solution of a particular max-min problem between the principal and nature. A natural question to ask is whether the optimal contract would remain optimal if the worst-case volatility process $\{ \nu_t \}$ were specified exogenously. Formally, this corresponds to what is known as a Bellman-Isaacs condition. As discussed in \cite{hansen2006robust}, this condition is important for the interpretation of the solution to a robust control problem. In particular, it allows for an ex-post Bayesian interpretation of the robust control problem. 

For the robust contracting problem described in this paper, the value function is in fact globally concave, and the optimal control of nature has no binding inequality constraints. One can verify (see \cite{fan1953minimax}, \cite{hansen2006robust}) that the Bellman-Isaacs condition is satisfied. Therefore, the optimal contract described in proposition \ref{main} is optimal in an ex-post Bayesian sense where the principal believes that volatility evolves according to \eqref{nu}, in a restricted space of contracts where changes in the agent's continuation payoff are locally linear in project cash flows. As such, it is reasonable to interpret my model as a model of endogenous belief formation about the volatility process.

\FloatBarrier

\subsection{Implementation and Asset Pricing Implications}
%Following \cite{biais2007dynamic} I show how to implement the optimal contract using cash reserves, debt, and equity\footnote{An alternative approach would be to follow \cite{demarzo2006optimal} and implement the optimal contract using a credit line, long-term debt, and equity. In this case it is trivial to show that the credit limit strictly increases and the level of long-term debt strictly decreases in the level of ambiguity aversion $1/\theta$. }

%%%%%%%%%%%%%%%%%%%%%%%%
%%% CREDIT LINE IMPLEMENTATION %%%
%%%%%%%%%%%%%%%%%%%%%%%%

\subsubsection{Credit Line Implementation}

Following \cite{demarzo2006optimal}, I show how to implement the optimal contract with a capital structure of equity, debt, and a credit line.
%\footnote{Alternatively, the optimal contract can be implemented using cash reserves, debt, and equity as in \cite{biais2007dynamic}}
The implementation is as follows:

\begin{itemize}
\item\emph{Equity:} The agent holds inside equity for a fraction $\lambda$ of the firm. Dividend payments are at the discretion of the agent. 
\item\emph{Long-term debt:} Long term debt is a consol bond that pays coupons at a rate $x = \mu -\frac{\gamma}{\lambda} \overline{W}$. If the firm ever defaults on a coupon payment, debt holders force liquidation. 
\item\emph{Credit line:} The firm has a revolving credit line with credit limit $C^L = \frac{\overline{W}}{\gamma}$. Balances on the credit line are subject to an interest rate $\gamma$. The firm borrows and repays funds on the credit line at the discretion of the agent. If the balance ever exceeds $C^L$, the project is terminated. 
\end{itemize}

The following proposition characterizes how this implementation changes with the level of ambiguity aversion. 
\begin{proposition}\label{securities}
As the level of ambiguity aversion $1/\theta$ increases
\begin{itemize}
\item The optimal credit limit strictly increases.
\item The face value of the optimal long-term debt strictly decreases.  
\end{itemize}
\end{proposition}
Note that the fraction of equity held by the agent is determined by the incentive compatibility constraints, and does not change with $\theta$. 

%Define the cash reserves $M_t$ as $\frac{1}{\lambda} W_t$. Then $M_t$ evolves according to
%\[
%d M_t = \gamma M_t dt + \sigma \nu^*(\lambda M_t) d B_t^{\nu^*} - \frac{1}{\lambda} d C_t^*
%\]
%under the worst-case probability measure, with initial condition $M_0 = W_0/\lambda$. Outside investors (i.e. the principal) hold debt with coupon payment $[\mu - (\gamma - r) M_t] dt$ and a fraction $1-\lambda$ of equity. The agent holds a fraction $\lambda$ of equity. Equity pays dividents $\frac{1}{\lambda} dC_t^*$. 

%\subsubsection{Asset Pricing Implications}
I consider asset prices in a representative agent setting where the principal is the representative investor who trades debt and equity, whereas the agent is an insider who is restricted from trading in either security. I take $r$ as the risk-free rate. Then I price securities under the worst-case belief measure of the principal. This approach is analogous to those taken in \cite{anderson2003quartet}, \cite{biais2007dynamic}, and \cite{miao2016robust}. \\

The value of equity is given by
\[
S_t = \mathbb{E}_t^{\nu^*} \left[ \int_t^\tau e^{-r(s-t)} \frac{1}{\lambda} d C_t^*  \right] 
\]
It is straightforward to obtain that the stock price is given by $S_t = S(W_t)$ where the function $S(\cdot )$ satisfies the ODE
\[
rS(W) = \gamma W S'(W) + \frac{1}{2} \lambda^2 \sigma^2 \nu^*(W)^2 S''(W)
\]
with boundary conditions $S(0)=0$ and $S'(\overline{W})=1$. A simple argument now shows that the equity premium is given by
\begin{equation}
\mathbb{E}_t \left[ \frac{dS_t}{S_t}  \right] - r = 
- \frac{1}{2} \lambda^2 \sigma^2 \left[ \nu^*(W_t)^2 - 1 \right]  \frac{S''(W_t)}{S(W_t)} 
\end{equation}
which is strictly positive in numerical computations. Note also that in the no-ambiguity benchmark, the equity premium is identically zero. 

Define the credit yield spread $\Delta_t$ by
\begin{equation}
\int_t^\infty e^{-(r+\Delta_t)(s-t)} ds = \mathbb{E}_t^{\nu^*} \left[  \int_t^\tau e^{-r(s-t)} ds  \right] 
\end{equation}
which when solved yields $\Delta_t = \frac{rT_t}{1-T_t}$ where $T_t = \mathbb{E}_t^{\nu^*}\left[e^{-r(\tau - t)}\right]$ is the time-$t$ price of one unit of consumption at the time of default. $T_t = T(W_t)$ satisfies the ODE
\[
r T(W) = \gamma W T'(W) + \frac{1}{2} \lambda^2 \sigma^2 \nu^*(W)^2 T''(W) 
\]
with boundary conditions $T(0)=1$ and $T'(\overline{W}) = 0$. 

%The value of the outstanding debt $D_t$ is given by
%\[
%D_t = \mathbb{E}_t^{Q^{\nu^*}} \left[  \int_t^\tau e^{-r(s-t)} [\mu - (\gamma - r) M_s] \ ds + e^{-r(\tau - t)} L   \right]
%\]
%Clearly $D_t = D(M_t)$ for some function $D(\cdot )$ satisfying the ODE
%\[
%rD = [ \mu - (\gamma - r)] + \gamma M D'(M) + \frac{1}{2} \sigma^2 \nu^*(\lambda M)^2 D''(M)
%\]
%with boundary conditions $D(0)=L$ and $D'(\overline{W}/\lambda) = 0$. 

\begin{figure}[h!]
\centering
\includegraphics[scale=0.5]{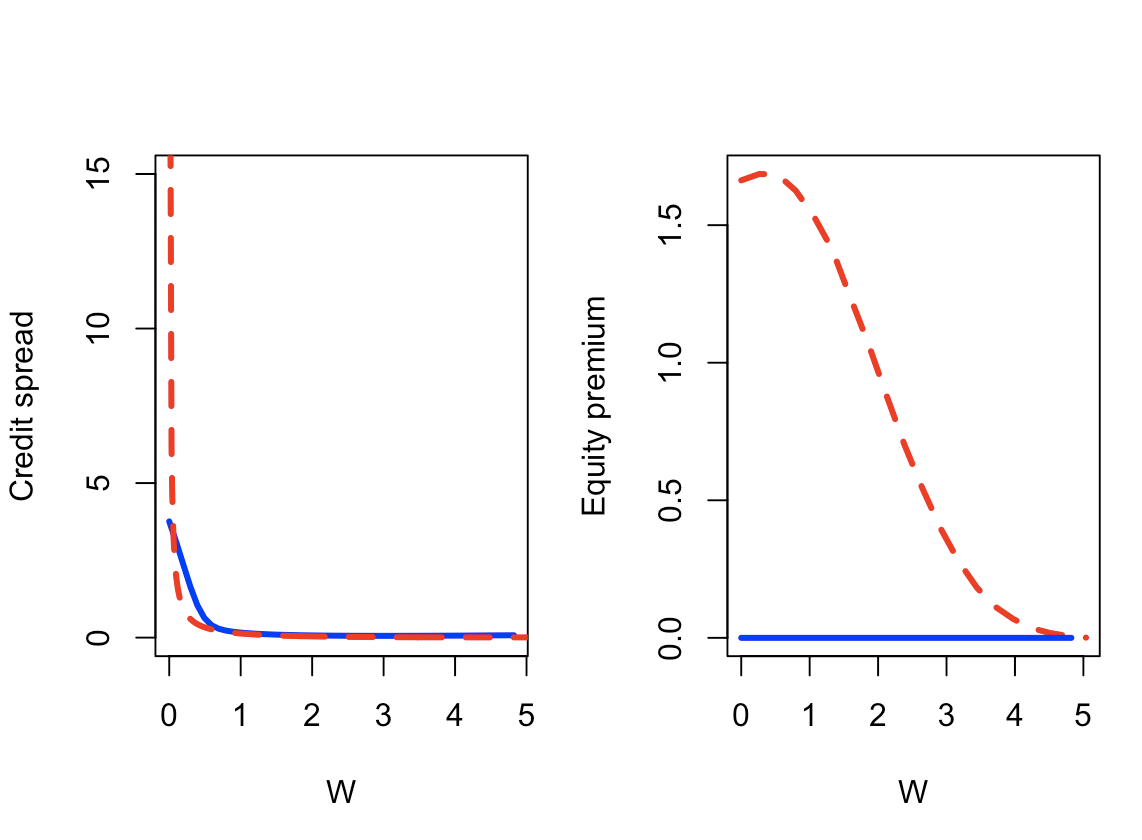}
\caption{Credit yield spread and equity premium as a function of $W$. Asset prices without ambiguity $(\theta = \infty)$ shown in blue. Asset prices with ambiguity $(\theta = 5)$ shown in red. }
\end{figure}

\FloatBarrier

%PLOTS!!!!
% Equity premia
% Credit yield spread
% maybe stock price, bond price (if you can figure out boundary conditions)

%%%%%%%%%%
%\section{Comparison with $G$-expectations}
%cite Peng, Epstein-Ji
%Complementary approach
%\[
%r F(W) = \sup_{c\geq 0, \phi} \inf_{\sigma \in [\underline{\sigma}, \overline{\sigma}]} \mu - c + (\gamma W - c) F'(W) + \frac{1}{2} \phi^2 \sigma^2 F''(W)
%\]
%Degenerate, \sigma = \overline{\sigma}, c = 0, \phi = \lambda
%worst-case volatility is constant => degenerate. 

\subsubsection{Cash-based implementation}
I briefly describe an alternate capital structure implementation of the optimal contract, similar to \cite{biais2007dynamic}, using equity, debt, and cash reserves. The firm holds cash reserves $M_t = \frac{W_t}{\lambda}$ which earn the risk-free interest rate $r$. The project payoffs $dY_t$ are put into the firm's cash account. Outside investors hold a fraction $1 - \lambda$ of equity, and debt which pays coupons at a state-dependent rate $[\mu - (\gamma - r) M_t] dt$, while the agent holds a fraction $\lambda$ of equity. Then the cash reserves evolve according to
\begin{equation}
\label{eq:cash}
dM_t = \underset{\text{interest}}{\underbrace{rM_t dt}}  + \underset{\text{project cash flows}}{\underbrace{dY_t}} - \underset{\text{dividends}}{\underbrace{\frac{1}{\lambda}  dC_t}} - \underset{\text{coupon}}{\underbrace{[ \mu - (\gamma - r)M_t] dt}}
\end{equation}
with $M_0 = W_0/\lambda$. One can easily verify that equation \eqref{eq:cash} agrees with the evolution for $W_t/\lambda$. Under the cash-based implementation, proposition \ref{payoffboundary} implies that higher levels of ambiguity aversion increase the amount of cash the firm will hold before it is willing to pay dividends. 

We see in both the credit line implementation and the cash-based implenentation that higher levels of ambiguity aversion increase the maximum ``financial slack'' that the firm is given under the optimal contract. In the credit line implementation this corresponds to a higher maximum credit limit, whereas in the cash-based implementation this corresponds to a higher cash buffer that the firm accumulates before paying dividends to equity holders. 

\subsection{The role of commitment}
The optimal contract described is proposition \ref{main} is not generically renegotiation-proof. For small values of $W$, the principal's value function $F$ under the optimal contract is increasing in $W$, so the principal and the agent can both be made better off by a one-off increase in the continuation value of the agent. To be renegotiation-proof, the principal's value function $F(W)$ must not have positive slope. 
However, it is possible to modify the contract described in proposition \ref{main} to obtain the optimal renegotiation-proof contract, which I describe in this section. Additionally, I show that the implied worst-case volatility under the optimal renegotiation-proof contract is strictly decreasing in the agent's continuation value. 

Renegotiation effectively raises the minimum payoff of the agent to a point $R$ such that $F'(R) = 0$. The agent's promised value evolves on the interval $[R, \overline{W}]$ according to
\begin{equation}
dW_t = \gamma W_t dt - dC_t  - \lambda \mu dt + \lambda \sigma \nu(W_t) d Z_t + d P_t
\end{equation}
where the processes $C$ and $P$ reflect $W_t$ at endpoints $\overline{W}$ and $R$ respectively. The project is terminated stochastically whenever $W_t$ is reflected at $R$. The probability that the project continues at time $t$ is
\begin{equation}
Pr(\tau \geq t) = \exp \left( - \frac{P_t}{R}  \right). 
\end{equation}
The optimal contract can still be implemented with equity, long-term debt and a credit line, though the level of long-term debt and the length of the credit line will be different.  

\begin{proposition}\label{nuincreasing}
Under the optimal renegotiation-proof contract, the worst-case volatility $\nu^*(W)^2$ is strictly increasing in $W$. 
\end{proposition}

The renegotiation-proof implementation contract is in a sense a more robust implementation than the implementation described in proposition \ref{main} in that it eliminates the incentive for the principal to renegotiate the contract with the agent. However, it still requires the principal to commit to a stochastic (unverifiable) liquidation policy. Without such commitment, there will generally be welfare loss to the principal. In particular, if the principal can only commit to deterministic liquidation policies, then the Pareto frontier is generally characterized by a solution to the same differential equation as before, but now with boundary conditions $F(0)=L$ and $F'(0) = 0$. Under this implementation, it is possible to show similar comparative statics as for the optimal contract with full commitment. 

\subsection{Comparison with alternative models}

\subsubsection{Comparison with $G$-expectations}
Consider the ``interval uncertainty'' or $G$-expectations formulation of ambiguity aversion. Assume that the adjustment cost function $\xi(\nu)$ faced by nature is given by
\begin{equation}\label{G}
 \xi(\nu) = \begin{cases} 0 & \text{ if } \nu \in [\underline{\sigma}/\sigma, \overline{\sigma}/\sigma ] \\
						\infty & \text{ otherwise. }
						\end{cases} 
						\end{equation}
This is equivalent to assuming that nature is free to choose any level of volatility $\sigma_t \in [\underline{\sigma},\overline{\sigma}]$ with no adjustment cost. This formulation of volatility ambiguity is precisely the $G$-expectation formulation of \cite{peng2007g}, and is similar to the $\kappa$-ignorance specification of \cite{chen2002ambiguity}. 

\begin{proposition}\label{Gcontract}
Consider the optimal contracting problem in which both the principal and the agent have interval uncertainty of the form \eqref{G}. Assume that $L < \frac{\mu}{r}$ and implementing high effort is optimal. Then the optimal contract is the same as that of the optimal contract without ambiguity aversion where both the principal and the agent believe the volatility level is $\overline{\sigma}$. 
\end{proposition}

\begin{proposition}\label{Gstatic}
The payoff boundary $\overline{W}$ of the optimal contracting problem with interval uncertainty is strictly increasing in $\overline{\sigma}$. 
\end{proposition}

\subsubsection{Comparison with drift ambiguity}

This paper is closely related to \cite{miao2016robust} who study a similar dynamic contracting problem where the principal is uncertain about the expected cash flows and is ambiguity-averse. They obtain similar asset pricing implications as I do; time-varying risk-premia that are generally higher for financially distressed firms. However, there are some key differences. Firstly, the optimal contracts are quite different. In my model, the incentive compatibility constraint always binds because the principal fears inefficient liquidation and therefore does not want the agent to bear any more risk than necessary. This preserves the optimality of the simple contractual form of \cite{demarzo2006optimal} and \cite{biais2007dynamic}. In their model however, the principal does not like drift ambiguity, and thus the the optimal contract will sometimes force the agent to bear more cash-flow sensitivity than necessary. As a result, the incentive compatibility constraint is an occasionally binding constraint, and their optimal contract is much more challenging to interpret. Second, the value function in my model is globally concave, so the Bellman-Isaacs condition holds. This means that it is valid to interpret my model as a model of endogenous belief formation. This is not the case in \cite{miao2016robust}. Thirdly, my model can accommodate ambiguity aversion on the part of the agent, without any reduction in the impact of ambiguity aversion. \cite{miao2016robust} do not model ambiguity aversion on the part of the agent, and in their framework, it would produce an offsetting effect which reduces the impact of ambiguity aversion on the optimal contract.

%%% comparison with risk-averse agent!!!

\subsection{Empirical implications}

Credit lines, also known as revolving credit facilities, are an extremely important form of firm financing. Empirically, credit lines account for more than a quarter of outstanding corporate debt of publicly traded firms and an even larger fraction for smaller, non-publicly traded firms.\footnote{See \cite{berger1995relationship}, \cite{sufi2007bank} and \cite{demarzo2006optimal}.} To the extent that smaller firms have more ambiguous riskiness of their cash flows, this is consistent with the predictions of my model. 

Under ambiguity aversion, the optimal contract may be interpreted as one that would emerge under a particular form of belief heterogeneity. As a reflection of the ambiguity aversion, the principal acts as if she believes that volatility is time-varying and strictly higher than the benchmark volatility. From the principal's point of view, the agent's beliefs appear to have too little uncertainty. This is consistent with  empirical evidence on managerial overconfidence as in \cite{landier2009financial} and \cite{ben2013managerial}. 

In terms of asset prices, my model predicts that the equity premium and credit yield spread are state-dependent and generally higher for firms closer to default. This is consistent with the literature on characteristic-based asset pricing (\cite{daniel1997evidence}, \cite{daniel1998characteristics}) as well as \cite{friewald2014cross} who find that firm's equity premium and credit spread are positively correlated.

\section{Conclusion}

This paper developed new preference formulations which capture ambiguity aversion towards unknown volatility. These \emph{moment-constrained variational preferences} were introduced in a static setting and their impact illustrated in a simple model of portfolio choice under quadratic utility. In this model, I showed how the degree of ambiguity aversion impacted the implied worst-case volatility as well as the optimal portfolio of the investor. I then derived a continuous-time limit in which ambiguity aversion towards unknown, potentially time-varying, volatility was not degenerate. These new continuous-time preferences were then compared with the $G$-expectations model of \cite{peng2007g}. The impact of these new continuous-time preferences was illustrated in a model of optimal security design under repeated moral hazard. I showed how the worst-case volatility of the principal depended on the distance to liquidation, whereas the agent always has full confidence in their benchmark model. I showed how ambiguity aversion increased the dividend ``hurdle'' in the optimal contract, and further showed how in a credit line implementation this corresponded to increasing the maximum draw on the credit line to the agent. Finally, I numerically illustrated some of the asset pricing implications of ambiguous volatility. 

For pedagogical simplicity and clarity, this paper focused exclusively on ambiguity aversion towards unknown volatility. However, there is no deep theoretical reason for this exclusivity. Future work could apply moment-constrained variational preferences to modelling uncertainty about other distributional features, and the volatility penalization could be used in conjunction with other approaches for modelling ambiguity aversion. 

The preference formulations developed in this paper can potentially be applied to a variety of other settings. One possibility is to examine their effect in a with moral hazard and endogenous investment, similar to \cite{demarzo2012dynamic} or \cite{bolton2013market}, and derive simultaneous implications for corporate investment and asset pricing. Another possibility would be to apply them to the problem of stress testing, where a bank regulator attempts to control the risk-taking of a bank without full confidence in a particular risk model. A third possibility would be to study a consumption-savings problem and investigate the impact of volatility ambiguity on precautionary savings.\footnote{I am grateful to an anonymous referee for this suggestion.} I leave these and other extensions to future research. 

%%% Need to say something about mathematical rigour, Peng, Nutz, etc.

\bibliography{ambiguous}

\clearpage

\appendix

\section{Proofs and derivations for section 2}
\subsection{Proofs for subsection \ref{subsec:scalar}}
\begin{lemma}
\label{a}
$s(\theta)$ can be expressed as $v(\theta) + \mu^2 + \sigma^2$ where $v(\theta)$ is the unique positive root of the cubic polynomial
\[
\frac{1}{2} \mu^2 C^2 \frac{1}{v + \mu^2 + \sigma^2} + \frac{\theta}{2} \frac{1}{\sigma^2} (v + \mu + \sigma^2) - \frac{\theta}{2} \log (v + \sigma^2) + D. 
\]
where
\begin{align*}
C &= (b - W_0 R_f) \\
D &= - \frac{1}{2} (b - W_0 R_f)^2 - \frac{\theta}{2}. 
\end{align*}
Additionally, $v(\theta)$ is strictly decreasing in $\theta$ and we have the limits $\lim_{\theta \to \infty} v(\theta) = 0$ and $\lim_{\theta \to 0} v(\theta) = \infty$. 
\end{lemma}

\subsection*{Proof of lemma \ref{a}}
Write $C = (b - W_0 R_f)$. Then the objective function in \eqref{smin} can be written as
\[
\frac{1}{2} \mu^2 C^2 \frac{1}{2} + \frac{\theta}{2} \frac{1}{\sigma^2} s - \frac{\theta}{2} \log (s - \mu^2) + D
\]
where $D$ is a constant that does not depend on $s$. We know that the minimizing $s$ is unique and satisfies $s > \mu^2 + \sigma^2$. Now, perform the following change-of-variables. Define $v = s - \mu^2 - \sigma^2$. The objective function, written now as a function of $v$ is given by
\[
\frac{1}{2} \mu^2 C^2 \frac{1}{v + \mu^2 + \sigma^2} + \frac{\theta}{2} \frac{1}{\sigma^2} (v + \mu + \sigma^2) - \frac{\theta}{2} \log (v + \sigma^2) + D
\]
Clearly the minimizing $v$ is unique and satisfies $v > 0$. Since the objective function is differentiable, the first-order condition gives a necessary condition that the optimal $v$ must satisfy. Note that this is not a sufficient condition unless there is a unique positive solution. The first-order condition for $v$ simplies to the following cubic polynomial equation for $v$,
\[
0 = \frac{\theta}{2} \frac{1}{\sigma^2} v^3 + \frac{\theta}{\sigma^2} (\mu^2 + \sigma^2) v^2 + \left[  \frac{\theta}{2 \sigma^2} (\mu^2 + \sigma^2)^2 - \frac{1}{2}\mu^2 C^2 \right] v - \frac{\mu^2 C^2}{2}. 
\]
To see that this equation has a unique positive solution, note that the terms of the cubic equation are in descending order, that the first two coefficients are positive and that the last is negative.\footnote{The third coefficient has ambiguous sign.} Hence by Descartes' rule of signs that the cubic equation has a unique solution. 

It follows from Topkis's theorem that $v(\theta)$ is decreasing in $\theta$. The limits follow immediately. \qed

Note that proposition \ref{s2} follows immediately from lemma \ref{a} and the observation that
\[
\nu^2(\theta) = \frac{v(\theta)}{\sigma^2} + 1. 
\]
Proposition \ref{s1} follows from lemma \ref{a} and the observation that 
\[
\phi^*(\theta) = \frac{1}{v(\theta) + \mu^2 + \sigma^2} \mu (b - W_0 R_f). 
\]

\subsection{Portfolio choice with ``multiplier'' preferences}
\label{appen:multiplier}

We consider the following portfolio choice problem with``unconstrained'' or ``multiplier'' preferences:
\begin{equation}
\sup_{\phi \in \mathbb{R}} \inf_{M \geq 0, \mathbb{E}_P[M] = 1} \mathbb{E}_P[ M U(\widetilde{W}) ] + \theta \Phi^u(M)
\end{equation}
where
\begin{align*}
\Phi^u(M) &= \mathbb{E}_P[M \log M]  \\
U(\widetilde{W}) &= - \frac{1}{2} (\widetilde{W} - b)^2 \\
\widetilde{W} &= W_0 R_f + \phi \widetilde{R} \\
\widetilde{R} &\overset{P}{\sim} \text{Normal}(\mu,\sigma^2). 
\end{align*}
As in the text, I assume that $\mu > 0$ and $b > W_0 R_f$. Then the solution to [whatever the equation number of the problem is] has the following properties:
\begin{itemize}
\item The minimizing $M$ implies a Normal distribution for $\widetilde{W}$. 
\item The optimal portfolio weight $\phi_u(\theta)$ is increasing in $\theta$. 
\item The worst-case mean $\mu_u(\theta)$ is increasing in $\theta$, and $\mu_u(\infty) = \mu$. 
\item The worst-case variance $\sigma^2_u(\theta)$ is increasing in $\theta$, and $\sigma_u^2(\infty) = \sigma^2$. 
\end{itemize}

I give the following argument: It follows from equation \eqref{mult} that the minimizing $M$ has an exponential tilting form. This implies that $\widetilde{W}$ will have a normal distribution under the change-of-measure induced by $M$, with distorted mean and variance $\widetilde{\mu}$ and $\widetilde{\sigma}^2$ respectively. $\Phi^u(M)$ can then be expressed in terms of $\widetilde{\mu}$ and $\widetilde{\sigma}^2$ as. 
\[
\Phi^u(M) = \frac{1}{2} \left[  \frac{\widetilde{\sigma}^2}{\sigma^2} + \frac{1}{\sigma^2} (\widetilde{\mu} - \mu)^2 - \log \left( \frac{\widetilde{\sigma}^2}{\sigma^2}  \right) - 1 \right] 
\]
Treating $\phi, \widetilde{\mu}$, and $\widetilde{\sigma}^2$ as fixed, we see that
\[
\mathbb{E}\left[ M U(\widetilde{W}); \phi \right] = - \frac{1}{2} \phi^2 (\widetilde{\mu}^2 + \widetilde{\sigma}^2) + \phi \widetilde{\mu} (b - W_0 R_f) - \frac{1}{2} (b - W_0 R_f)^2
\]
Maximizing over $\phi$, we see that
\[
\phi(M) = \frac{1}{\widetilde{\mu}^2 + \widetilde{\sigma}^2} \widetilde{\mu} (b - W_0R_f)
\]
Now, substituting in the optimized value of $\phi(M)$ we obtain
\[
\mathbb{E}\left[ M U(\widetilde{W})\right] = \frac{1}{2} \frac{\widetilde{\mu}^2}{\widetilde{\mu}^2 + \widetilde{\sigma}^2} (b - W_0R_f)^2 - \frac{1}{2} (b - W_0 R_f)^2.
\]
Define
\begin{align*}
L(\widetilde{\mu},\widetilde{\sigma}^2;\theta) = \frac{1}{2} \frac{\widetilde{\mu}^2}{\widetilde{\mu}^2 + \widetilde{\sigma}^2} (b - W_0R_f)^2 - \frac{1}{2} (b - W_0 R_f)^2 + \frac{\theta}{2} \left[ \frac{\widetilde{\sigma}^2}{\sigma^2} + \frac{1}{\sigma^2} (\widetilde{\mu} - \mu)^2 - \log \left( \frac{\widetilde{\sigma}^2}{\sigma^2}  \right) - 1 \right]. 
\end{align*}
Observe that $L(\widetilde{\mu},\widetilde{\sigma}^2;\theta) = \max_{\phi} \mathbb{E}\left[ M U(\widetilde{W})\right] + \theta \Phi^u(M)$ when $M$ is restricted to imply that $\widetilde{W} \sim \text{Normal}(\widetilde{\mu},\widetilde{\sigma}^2)$. The solution for $\mu_u(\theta)$ and $\sigma^2_u(\theta)$ can thus be obtained by solving 
\begin{equation}
\min_{\widetilde{\mu},\widetilde{\sigma}^2 \geq 0} L(\widetilde{\mu},\widetilde{\sigma}^2;\theta). 
\end{equation}
It follows directly from Topkis' monotonicity theorem that $\mu_u(\theta)$ is strictly increasing in $\theta$, or equivalently, strictly decreasing in $1/\theta$. Since $\mu_u(\infty) = \mu$, we see that $\mu_u(\theta) < \mu$. 

The first-order condition for $\sigma^2_u(\theta)$ implies that
\[
\frac{1}{2} \frac{\mu_u(\theta)^2}{\mu_u(\theta)^2 + \sigma^2_u(\theta)} (b - W_0 R_f) = \frac{\theta}{2} \left[ \frac{1}{\sigma^2} - \frac{1}{\sigma^2_u(\theta)} \right]
\]
from which we see that $\widetilde{\sigma}^2$ must be strictly greater that $\sigma^2$. It follows from Topkis' theorem that $\sigma^2_u(\theta)$ is strictly increasing in $\theta$.

\section{Proofs for section 4}

\subsection*{Proof of proposition \ref{main}}

Note that $F(W)$ is necessarily a concave solution to the HJBI equation
\[
r F(W) = \sup_{c \geq 0, \phi \geq \lambda} \ \inf_{\nu \geq 0} \ \mu - c + \xi(\nu) + (\gamma W - c) F'(W) + \frac{1}{2} \phi^2 \sigma^2 \nu^2 F''(W)
\]
on $(0,\overline{W})$ with optimal controls $c = 0$, $\phi = \lambda$, and $\nu^2 = \nu^*(W)^2$. It follows that for any $\phi \geq \lambda$ we have
\[
\inf_{\nu \geq 0} \ \mu  + \xi(\nu) + \gamma F'(W) + \frac{1}{2} \phi^2 \sigma^2 \nu^2 F''(W) - r F(W) \leq 0
\]

Define 
\[
G_t = \int_0^t e^{-rs} \left( dY_s - dC_s + \xi(\nu_s) ds  \right) + e^{-rt} F(W_t)
\]
Then note that
\begin{align*}
e^{rt}  dG_t = \left( \mu + \xi(\nu_t) + \gamma W_t F'(W_t) + \frac{1}{2} \phi_t^2 \sigma^2 \nu_t^2 F''(W_t) - rW_t  \right) dt \\
- (1 + F'(W_t)) dC_t + (1+\phi_t F'(W_t)) \sigma \nu_t dZ_t
\end{align*}
Note that $F'(W_t) \geq -1$ so that $-(1+F'(W_t)) \leq 0$. Additionally, under the worst-case $\nu_t$ we have
\[
\mu + \xi(\nu_t) + \gamma W_t F'(W_t) + \frac{1}{2} \phi_t^2 \sigma^2 \nu_t^2 F''(W_t) - rW_t \leq 0. 
\]
Thus $G_t$ is a supermartingale. It is a martingale only if $\phi_t = \lambda, W_t \leq \overline{W}$ for $t \geq 0$ and $C_t$ is increasing only when $W_t \geq \overline{W}$. 

Now, we can bound the principal's time-0 payoff for an arbitrary incentive compatible contract. Note that $F(W_\tau) = L$. We have
\begin{align*}
&\inf_{\nu \geq 0} \mathbb{E} \left[  \int_0^\tau e^{-rs} \left\{ dY_s - dC_s + \xi (\nu_s) ds  \right\}  + e^{-r\tau} L  \right] \\
&= \int_{\nu \geq 0} \mathbb{E} \left[ G_{t \wedge \tau} + \mathbf{1}_{t \leq \tau} \left( e^{-rs} \left\{ dY_s - dC_s + \xi(\nu_s) ds \right\} + e^{-r\tau} L - e^{-rt} F(W_t)   \right) \right] \\
&\leq \underset{\leq G_0 = F(W_0)}{\underbrace{\mathbb{E} \left[ G_{t \wedge \tau} \right]}}
+ e^{-rt} \inf_{\nu \geq 0} \mathbb{E} \left[ \mathbf{1}_{t \leq \tau} \underset{\leq \mu/r - W_t}{\underbrace{\mathbb{E}_t \left[ \int_t^\tau e^{-r(s-t)} \left\{ dY_s - dC_s + \xi(\nu_s) ds \right\} + e^{-r(\tau-t)} L\right]}} - F(W_t)   \right] 
\end{align*}
where the second inequality follows from the first-best bound. Since $F'(W) \geq -1$ we have $\mu/r - W - F(W) \leq \mu/r - L$. Letting $t \to \infty$ we see that
\[
\inf_{\nu \geq 0} \mathbb{E} \left[ \int_0^\tau e^{-rs} \left\{ dY_s - dC_s + \xi(\nu_s) ds \right\} + e^{-r\tau} L  \right] \leq F(W_0). 
\]
\qed

\subsection*{Proof of proposition \ref{welfare} }
Applying Dynkin's formula to write the value function as an integral of the differential generator and then differentiating under the integral sign and applying the envelope theorem gives
\begin{align*}
\frac{\partial}{\partial \theta} F(W) = \mathbb{E} \left[ \int_0^\tau e^{-rt} \frac{1}{2} \left( \nu^*(W_t)^2 - 1 - \log(\nu^*(W_t)^2)  \right)\ dt \bigg| W_0 = W    \right] > 0. 
\end{align*}
\qed

\subsection*{Proof of proposition \ref{payoffboundary}}
Differentiate the boundary condition $r F(\overline{W}) + \gamma \overline{W} = \mu$ and use the smooth pasting condition $F'(\overline{W}) = -1$ to obtain
\[
r \left[ \frac{\partial}{\partial \theta} F(\overline{W}) - \frac{\partial \overline{W}}{\partial \theta}    \right] + \gamma \frac{\partial \overline{W}}{\partial \theta} = 0
\]
which gives
\[
\frac{\partial \overline{W}}{\partial \theta} = - \frac{r}{\gamma - r} \frac{\partial}{\partial \theta} F(\overline{W}) < 0. 
\]
\qed

\subsection*{Proof of proposition \ref{agent}}
Let $h(W)$ denote the agent's value function under the contract described in proposition \ref{main}. The HJBI equation for the agent is given by
\[
\gamma h(W) = \lambda \mu (1-a) + h'(W) (\gamma W + \lambda \mu (a-1) ) + \frac{1}{2} \lambda^2 \sigma^2 \nu^2 h''(W) +
\frac{\tilde \theta}{2} \left\{ \nu^2 - 1 - \log \nu^2 \right\}
\]
on $[0,\overline{W}]$ with boundary conditions $h(0)=0$ and $h'(\overline{W}) = 1$. Now, guess and verify that $h(W) = W$ is a solution with optimal controls $\nu(W)=1$ and $a(W) = 1$. It is easy to show that this solution must be unique. 
\qed

\subsection*{Proof of proposition \ref{securities}}
This follows immediately from proposition \ref{payoffboundary}
\qed

\subsection*{Proof of proposition \ref{nuincreasing}}
Differentiating \eqref{ode} w.r.t. $W$ we obtain
\[
0 = (\gamma - r) F'(W) + \gamma W F''(W) + \frac{\theta}{2} \frac{ \lambda^2 \sigma^2 F'''(W)}{\theta + \lambda^2 \sigma^2 F''(W)} 
\]
Note that the first term is negative since $\gamma > r$ and $F'(W) < 0$ on the interval $(R,\overline{W}]$. The second term is negative since $F''(W) < 0$ for $W < \overline{W}$. Thus the third term must be strictly positive. This can only happen if $F'''(W)$ is strictly positive. The result now follows from \eqref{nu}. 
\qed

\subsection*{Proof of proposition \ref{Gstatic}}
This follows immediately from proposition \ref{Gcontract} and Appendix B of \cite{demarzo2006optimal}
\qed

\end{document}